\documentclass[10pt, conference, compsocconf]{IEEEtran}
\IEEEoverridecommandlockouts
\usepackage{cite}
\usepackage{amsmath,amssymb,amsfonts}
\usepackage{algorithm}
\usepackage{algpseudocode}
\usepackage{graphicx}
\usepackage{textcomp}
\usepackage{xcolor}
\usepackage{microtype}
\def\BibTeX{{\rm B\kern-.05em{\sc i\kern-.025em b}\kern-.08em
    T\kern-.1667em\lower.7ex\hbox{E}\kern-.125emX}}

\usepackage{bm}
\usepackage{braket}
\usepackage{cuted}
\usepackage{subcaption}
\usepackage{widetext}
\usepackage[hyphens]{url}
\usepackage[colorinlistoftodos]{todonotes}

\graphicspath{{./fig/}}

\newcommand{\bYi}{\ensuremath{\bar{Y_i}}}

\newcommand{\pmi}{\ensuremath{p^{m_i}}}
\newcommand{\qmi}{\ensuremath{q^{m_i}}}
\newcommand{\stdseq}{\ensuremath{\sigma_{\mathrm{seq}(i)}}}
\newcommand{\stdshot}{\ensuremath{\sigma_{\mathrm{shot}(i)}}}

\newcommand{\eqnref}[1]{(\ref{#1})}

\begin{document}

\title{Sampling Strategy Optimization for \\
Randomized Benchmarking
}

\author{\IEEEauthorblockN{Toshinari Itoko}
\IEEEauthorblockA{\textit{IBM Quantum} \\
\textit{IBM Research - Tokyo}\\
Tokyo, Japan \\
itoko@jp.ibm.com}
 \and
\IEEEauthorblockN{Rudy Raymond}
\IEEEauthorblockA{\textit{IBM Quantum} \\
\textit{IBM Research - Tokyo}\\
Tokyo, Japan \\
rudyhar@jp.ibm.com}
}

\maketitle

\begin{abstract}
Randomized benchmarking (RB) is a widely used method for estimating the average fidelity of gates implemented on a quantum computing device.
The stochastic error of the average gate fidelity estimated by RB depends on the sampling strategy
(i.e., how to sample sequences to be run in the protocol).
The sampling strategy is determined by a set of configurable parameters (an RB configuration) that includes
Clifford lengths (a list of the number of independent Clifford gates in a sequence) and
the number of sequences for each Clifford length.
The RB configuration is often chosen heuristically and there has been little research on its best configuration.
Therefore, we propose a method for fully optimizing an RB configuration so that the confidence interval of the estimated fidelity is minimized while not increasing the total execution time of sequences.
By experiments on real devices, we demonstrate the efficacy of the optimization method against heuristic selection in reducing the variance of the estimated fidelity.
\end{abstract}


\section{Introduction} \label{sec:introduction}
As more and larger quantum computers without fault tolerance are physically implemented,
there is a growing need for methods to benchmark their performance. Straightforward tomography-based methods require 
measurements that scale exponentially with the number of qubits~\cite{chuang1997prescription} and thus are not scalable. Utilizing advanced techniques, such as, 
compressed sensing~\cite{gross2010quantum} and samplings~\cite{flammia2011direct,da2011practical}, the scaling bottleneck 
can be avoided to obtain the average gate fidelity but such methods are not robust against state preparation and measurement (SPAM) errors. 
Randomized benchmarking (RB)~\cite{knill2008randomized,magesan2011scalable} is an efficient and robust method and widely used in practice 
for estimating the average fidelity of a gate set implemented on a quantum computing device~\cite{barends2014superconducting,muhonen2015quantifying,chow2009randomized,corcoles2013process,mckay2017efficient,mckay2019three}.
%
For example, IBM Quantum systems~\cite{ibmquantum} report their 1-qubit and 2-qubit gate error rates
calculated from the estimated average gate fidelity via RB.
Hence it is important to minimize the stochastic error in the estimated fidelity by RB
so that we can sufficiently track the drift in the gate fidelity over time,
which reflects imperfection of controlling physical devices.

The standard RB is a protocol composed of three steps.
First, it generates sets of sequences with random Clifford gates such that all sequences in each set have the same Clifford length (the number of independent Clifford gates in a sequence) but the length varies from set to set.
Then, it executes the sequences to measure the survival rate (i.e., probability of observing the initial state) for each Clifford length.
Finally, it estimates the exponential decay rate, which can be linearly transformed to the average gate fidelity, from the survival rate data.
The protocol has a three-fold sampling structure:
sampling Clifford lengths at the estimation (fitting) step,
sampling random sequences at the sequences generation step, and
sampling bit strings at the sequences execution step.
Therefore, the estimated decay rate (or equivalently, the average gate fidelity) is intrinsically subject to stochastic errors that depend on the sampling strategy.
The sampling strategy is determined by a set of configurable parameters that defines how to sample sequences to be run
(we call it an RB configuration),
e.g., Clifford lengths and the number of sequences for each Clifford length.
%

There are several studies that partially address the problem of finding an optimal RB configuration.
The number of sequences at each Clifford length that achieves a desired confidence level was loosely estimated by using Hoeffding bound~\cite{magesan2012characterizing}.
A tighter estimation comparable to the number used in practice was provided, assuming to use the ordinary least squares estimator in the fitting step~\cite{epstein2014investigating}.
It was suggested that varying the number of sequences depending on Clifford length may improve the reliability of estimated decay rate~\cite{helsen2019multiqubit}.
Finding the best maximum Clifford length was also discussed in~\cite{harper2019statistical}.
However, none of them addressed the problem of both optimizing Clifford lengths and the number of sequences at the same time.

\begin{figure*}[htbp]
\centerline{\includegraphics[clip, width=\textwidth]{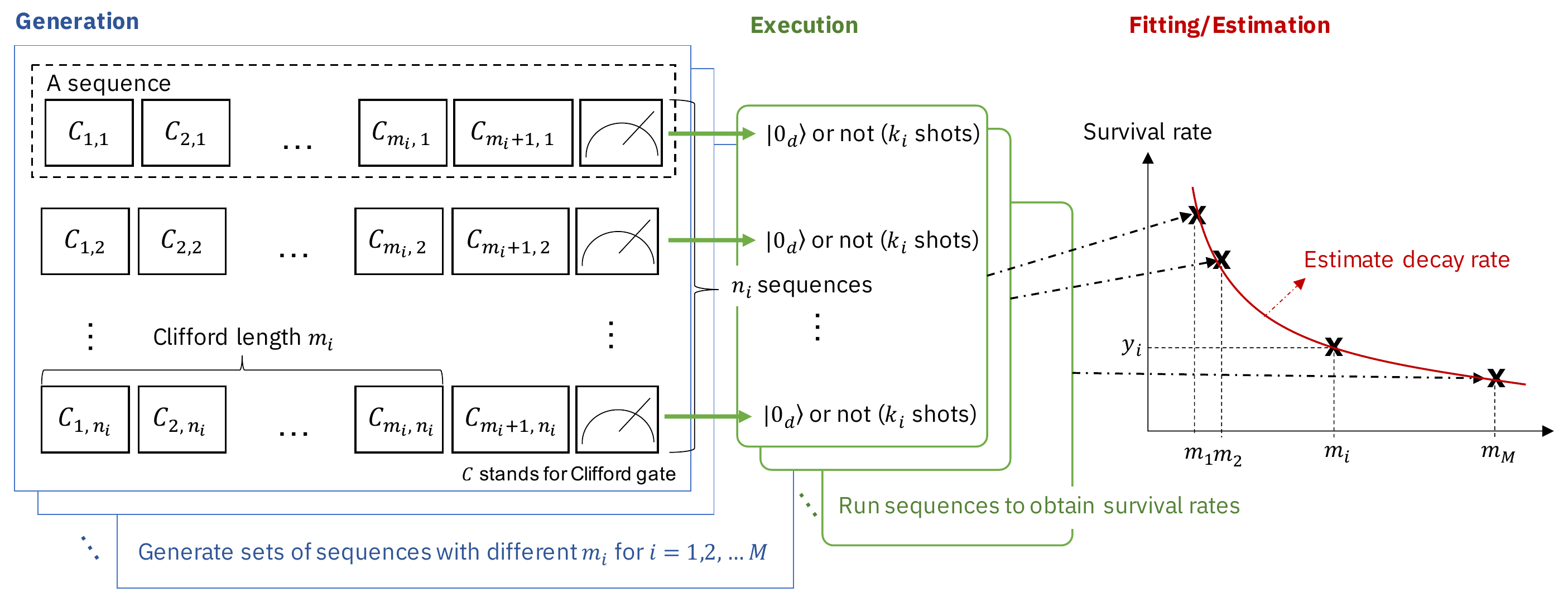}}
\caption{The standard RB (randomized benchmarking) protocol}
\label{fig:rb}
\end{figure*}

In this paper, we provide a method for finding an optimal sampling strategy (RB configuration)
that includes both Clifford lengths and the number of sequences for each Clifford length.
The optimal strategy yields a minimal confidence interval of the estimated average gate fidelity
within a given time budget for running the sequences (Section~\ref{sec:method}).
To make it possible, we customize techniques for analyzing statistical errors of estimated values in RB.
First, we construct a simple model for the variance of average survival rate (Section~\ref{sec:var-model})
in order to explain the variance data sampled from real devices,
which may vary depending on Clifford length.
Based on the variance model, we derive an explicit expression that approximates the confidence interval of the estimated decay rate as a function of the RB configuration (Section~\ref{sec:conf-interval}).
The derivation enables us to formulate the optimization of the RB configuration as a mathematical optimization problem (Section~\ref{sec:formulation}).
We also experiment on real devices to show that our method can find a better configuration
that achieves smaller variance in the resulting estimated decay rate
than typical heuristic configurations in practice (Section~\ref{sec:experiment}).

\section{Preliminaries}

We first set some notation and briefly review the standard RB protocol.
We suppose to benchmark a $d$-qubit quantum system,
which can be represented by the $D=2^d$ dimensional Hilbert space.
We assume the initial state is always set to $\ket{0_d}$ and
the measurement is a projection onto the computational basis, i.e. $\ket{0}$ or $\ket{1}$ for each qubit.

The standard RB protocol~\cite{magesan2011scalable,magesan2012characterizing} is described as follows.
\begin{description}
\item[Step 1] (Generation) Create $n_i$ sequences, each of which is a sequence of $m_i+1$ Clifford gates followed by a measurement as shown in Fig.~\ref{fig:rb}.
The first $m_i$ gates are chosen uniformly at random from the $d$-qubit Clifford group
and the $(m_i+1)$-th gate is uniquely determined as the inverse of the composition of the first $m_i$ gates.
\item[Step 2] (Execution) Run the $n_i$ sequences, measure the survival rate (i.e., the number of times $\ket{0_d}$ is observed divided by $k_i$ trials)
for each sequence, and average over the $n_i$ survival rates to
obtain the average survival rate $y_i$.
\item[Step 3] (Fitting/Estimation) Repeat Step 1 and 2 for different Clifford lengths $[m_1, m_2, \ldots, m_M]$
and then fit the results $[y_1, y_2, \ldots, y_M]$ to the decay model (e.g. $y_i \sim a\,\pmi + b$)
to estimate the decay rate $p$, which provides the average gate fidelity
$F_{\mathrm{avg}}=p + \frac{1-p}{D}$.
\end{description}
Here, the Clifford group is defined as the normalizer of the Pauli group.
The size of the Clifford group grows superexponentially with the number of qubits $d$,
e.g., $24$ (when $d=1$), $11,520$  (when $d=2$), and $92,897,280$ (when $d=3$)\cite{ozols2008clifford}.
The inverse of the composition of Clifford gates is efficiently (in time polynomial in $d$)
computable on a classical computer
thanks to the tableau representation of $d$-qubit Clifford group operations,
e.g.~\cite{dehaene2003clifford,aaronson2004improved},
so the final gate in any sequence is as well.
Each element of the group can be
generated by elementary gates, e.g. the phase, Hadamard and controlled-NOT (CNOT) gates~\cite{aaronson2004improved,bravyi2021clifford}.

The standard RB protocol contains the following three types of parameter sets
that determines its sampling strategy:
\begin{itemize}
\item Clifford lengths: $\bm{m} = [m_1, m_2, \ldots, m_M]$.
A \emph{Clifford length} means the number of independent Clifford operations in a sequence.
\item List of the number of sequences: $\bm{n} = [n_1, n_2, \ldots, n_M]$.
The number of sequences with Clifford length $m_i$ is denoted by $n_i$.
\item List of the number of shots: $\bm{k} = [k_1, k_2, \ldots, k_M]$.
A \emph{shot} means a single execution of a sequence.
The number of shots for each sequence with Clifford length $m_i$ is denoted by $k_i$,
which is often fixed to a common constant $k$.
\end{itemize}
We refer to these sets of parameters as \emph{an RB configuration}.

We denote a function $f$ of variable $x$ with parameter $\theta$ by $f(x; \theta)$.
We consider the simplest decay model
$f(m; p, a, b) = a \, p^m + b$
as a function to be fitted in the estimation step of RB.
Here $p$ represents the decay rate.
The coefficients $a$ and $b$ absorb the state preparation and measurement (SPAM) errors
as well as the error on the final gate.
If there were no such errors, $a=1-\frac{1}{D}$ and $b=\frac{1}{D}$ would hold
(see e.g. \cite{magesan2012characterizing} for detailed analysis of the decay model).

We denote the average survival rate over $n_i$ random sequences
(and $k_i$ shots for each sequence) with $m_i$ Cliffords by $\bYi$
and the standard deviation of $\bYi$ by $\sigma_{\bYi}$, respectively.
Because $\sigma_{\bYi}$ depends on an RB configuration ($m_i$, $n_i$, $k_i$),
it may be denoted by $\sigma_{\bYi}(m_i, n_i, k_i)$ explicitly.

\section{Method} \label{sec:method}

\begin{figure*}[htbp]
\centerline{\includegraphics[clip, width=\textwidth]{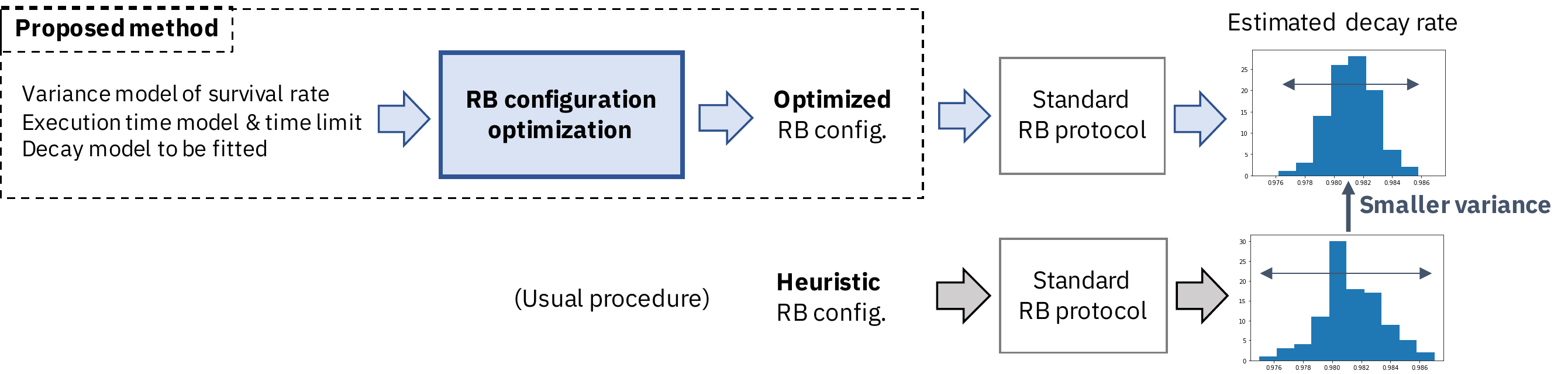}}
\caption{Overview of the proposed method for RB configuration optimization}
\label{fig:overview}
\end{figure*}

\subsection{Overview of RB Configuration Optimization}
We provide an overview of our method for optimizing an RB configuration in Fig.~\ref{fig:overview}.
The three main features of our method are as follows.
\begin{itemize}
\item It requires a variance model of the survival rate and an execution time model in addition to the decay model.
It also requires prior estimates of some parameters used in those models.
\item It derives the confidence interval of the estimated decay rate as a function of an RB configuration, assuming the \emph{weighted least squares} (WLS) estimator is used in the fitting step of the RB.
\item It formulates a problem of minimizing the confidence interval under the constraint that the predicted execution time of sequences must be within a given time budget. 
Solving the optimization problem determines an optimal RB configuration.
\end{itemize}
Our method can be seen as preprocessing of the standard RB protocol.
Once an optimal configuration is computed by our method,
we only run the standard RB protocol following the configuration,
and we obtain the estimated decay rate with minimal sampling errors.

In the following sections, we will detail our method.
First, we introduce the variance model
given that the variance of the survival rate differs depending on Clifford length (Section~\ref{sec:var-model}).
To take the heteroskedasticity into account,
we use the WLS estimator in the fitting.
We derive the confidence interval of the estimated decay rate with the WLS estimator
as a function of an RB configuration and
use it as the objective function to be minimized  (Section~\ref{sec:conf-interval}).
We constrain the total execution time to avoid
increasing the number of samples (e.g. the number of sequences)
infinitely to decrease the sampling error.
For that, we introduce a simple model to predict execution time (Section~\ref{sec:time-model}).
Finally, we provide a formulation for the problem of optimizing RB configuration as a mathematical optimization problem (Section~\ref{sec:formulation}).

\subsection{Variance Model of Survival Rate} \label{sec:var-model}

Although our method can accept any model of the variance of the average survival rate,
models with fewer parameters are preferable
because the method requires prior estimates of the parameters.
The goal of the variance model is to capture how the variance of the survival rate varies depending on Clifford length.
Any model of upper, average, or lower bound is acceptable for this purpose provided that
it approximates the form of the function over the entire region of Clifford length.
However, to the best of our knowledge, there is no model perfectly suitable for the use in RB configuration optimization.
For example, the upper bound in~\cite{wallman2014randomized} only provided analysis of the variance for a specific region of Clifford length.
An improved bound without such a restriction on Clifford length was proposed in~\cite{helsen2019multiqubit} for a variant of the standard RB,
however, how well the bound fits to real-world variance data was not discussed.

We model the variance of the average survival rate $\sigma_{\bYi}^2$ as follows.
First, we assume the sequence sampling error is independent of the shot sampling error,
and the total variance is given as the sum of those variances:
\begin{equation} \label{eqn:varYmodel}
\sigma_{\bYi}^2 = \frac{1}{n_i} \{ \stdseq^2 + \stdshot^2 \}
\end{equation}
Then we approximate each of them by, respectively, 
\begin{equation} \label{eqn:varcirc}
\stdseq^2 \approx \beta \, \qmi \, (1 - \qmi),
\end{equation}
\begin{equation} \label{eqn:varshot}
\stdshot^2 \approx \frac{\mu_i \, (1 - \mu_i)}{k_i}
\mbox{ with }
\mu_i = \left(1 - \frac{1}{D} \right) \, \hat{p}^{m_i} + \frac{1}{D}.
\end{equation}
Note that we expect the parameter $q$ should be close to the decay rate $p$.
Here, the approximate shot sampling error~\eqnref{eqn:varshot} is derived
under the strong assumptions that the mean of the survival rate for all the sequences with a Clifford length $m_i$ could be the same $\mu_i$ and that there are no errors in SPAM or the final Clifford operation (see Appendix~\ref{app:var-model} for the details).
The approximate sequence sampling error~\eqnref{eqn:varcirc} is empirical,
but it can be roughly explained by the effect of gate-dependent errors.
As Clifford length increases, the variance once increases as the tail of the distribution of the survival rate widens due to the increased variations of Clifford gates in a sequence.
However, the variance eventually converges to zero as Clifford length $m_i$ approaches infinity
because the survival rate is bounded within 0 to 1 and
its average decays exponentially as $m_i$ grows.

In fact, this model explains real-world variance data very well (see Fig.~\ref{fig:varYfit})
even though it has no strong theoretical justification
(see Appendix~\ref{app:var-model} for more examples).
\begin{figure}[htbp]
\centering
\includegraphics[width=.95\columnwidth]{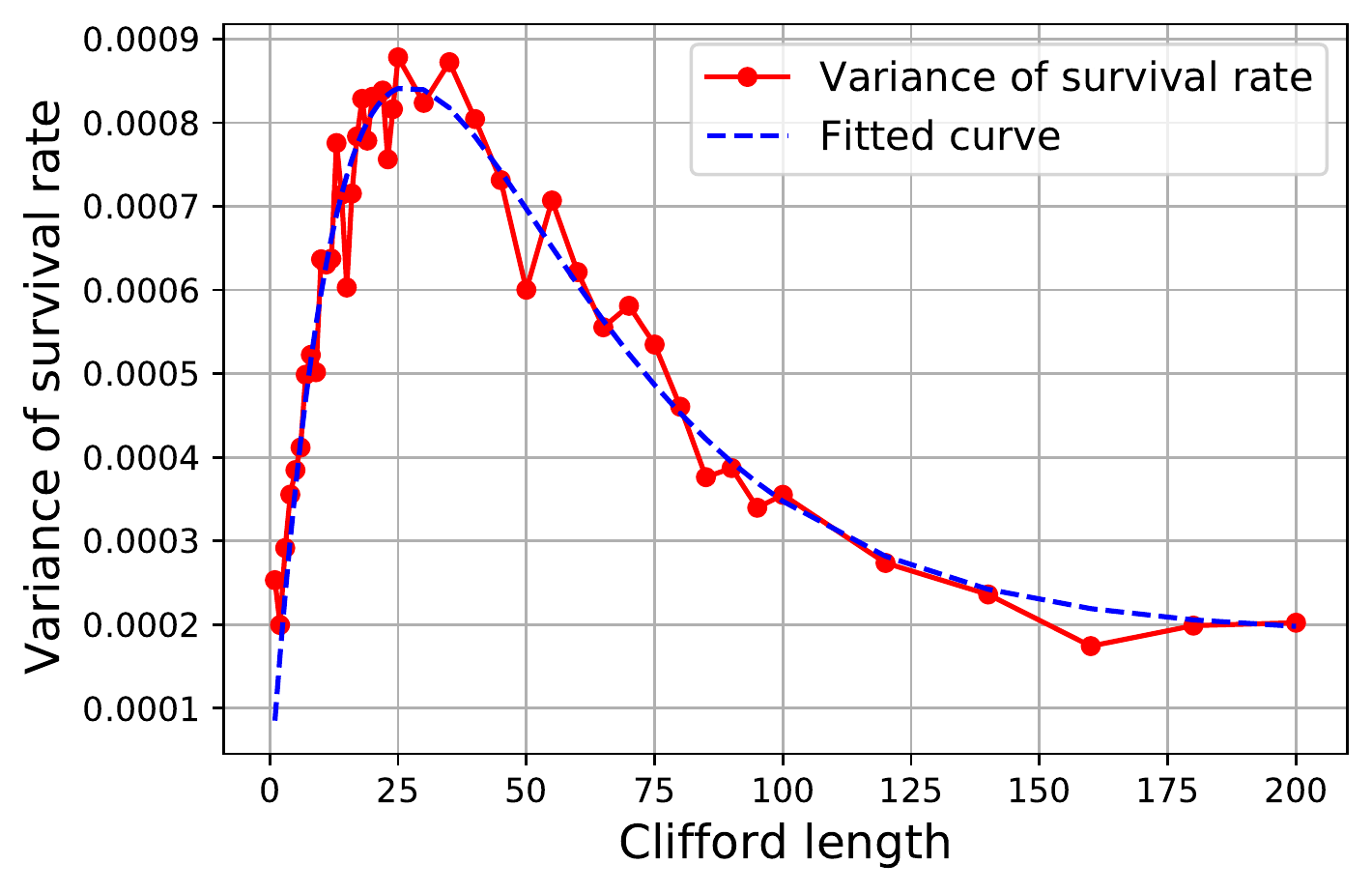}
\caption{Sample variance of survival rate for each Clifford length on a real device and their fitted curve to the variance model}
\label{fig:varYfit}
\end{figure}
In the figure, the sample variance of the survival rate for each Clifford length is plotted as red points,
and the blue dashed line shows the fitted curve to the model $n_i \, \sigma_{\bYi}^2$
defined by~\eqnref{eqn:varYmodel}--\eqnref{eqn:varshot}.
The survival rates for each Clifford length are sampled by running 400 sequences (each with 1,000 shots) on an IBM Quantum system \textsf{ibmq\_athens}.

As also shown in Fig.~\ref{fig:varYfit},
the variance of survival rate is not uniform regarding Clifford length in general.
That suggests the assumption of uniform variance,
which is assumed in the ordinary least squares (OLS) estimator, is not satisfied
as mentioned in several previous studies, e.g.~\cite{granade2015accelerated,helsen2019multiqubit}.
To take the heteroskedasticity into accounts,
using the iteratively reweighted least squares estimator was recommended in~\cite{helsen2019multiqubit}
and a Bayesian-based estimation algorithm was proposed in~\cite{granade2015accelerated}.
In this paper, we assume to use the weighted least squares (WLS) estimator in the fitting.
Recall that we are developing a method for optimizing the RB configuration,
not comparing the performance of other estimators with the OLS or WLS estimator.
The WLS estimator is suitable for our purpose because
it enables us to analytically derive the confidence interval of the estimated decay rate
as shown in the next section.

\subsection{Confidence Interval of Estimated Decay Rate} \label{sec:conf-interval}
The confidence interval of the decay rate estimated by the ordinary least squares (OLS) estimator
is explicitly given in \cite{epstein2014investigating}.
We extend their analysis to the case of the weighted least squares (WLS) estimator.

The $(1-\alpha)$ confidence interval of decay rate $p$
at the estimate $\hat{p}$ of a nonlinear regression model
$y_i = f(m_i; p, a, b) = a\, p^{m_i} +b$
by the WLS estimator with a weight matrix $W = diag(w_1, w_2, \ldots, w_n)$
is approximately given by
\begin{equation} \label{eqn:ci}
|p - \hat{p}| \leq t_{M-3, 1-\alpha/2} \, \sqrt{H \, s^2}.
\end{equation}
This is obtained by replacing $y_i$ with $\sqrt{w_i}\,y_i$ and $f(m_i)$ with $\sqrt{w_i}\,f(m_i)$ in the confidence interval of the estimated decay rate by the OLS estimator.
In the above~\eqnref{eqn:ci},
$t_{M-3, 1-\alpha/2}$ is the $(1-\frac{\alpha}{2})$ percentile of the $t$-distribution with $M-3$ degrees of freedom,
$s^2$ is the weighted sample average of the squared residuals, i.e.
\begin{equation} \label{eqn:s2}
s^2 = \frac{\sum_{i=1}^{M}{w_i \left| y_i - (\hat{a} \, \hat{p}^{m_i} + \hat{b}) \right|^2}}{M-3},
\end{equation}
and $H$ is the scaling factor that reflects how far $s^2$ extends in the $\hat{p}$ axis,
which is defined as
\begin{equation} \label{eqn:bigH}
H \equiv
\left[\left(J(\bm{\hat{\theta}})^T W J(\bm{\hat{\theta}}) \right)^{-1}\right]_{\hat{p}\hat{p}}
\end{equation}
Here $J(\bm{\hat{\theta}})$ is the Jacobian matrix of $f(m_i; \bm{\theta})$ ($i=1, \ldots, M$)
at $\bm{\theta}=\bm{\hat{\theta}}$ with $\bm{\theta}=[p, a, b]$, whose $i$-th row is given by
\begin{equation}
J_{i, *}(\bm{\hat{\theta}}) = \left[
\hat{a} \, m_i \, \hat{p}^{m_i-1},\ 
\hat{p}^{m_i},\ 
1
\right]
\end{equation}
Note that $H$ is an element of the inverse of the $3\times3$ matrix; hence, it can be analytically computed
(see Appendix~\ref{app:conf-interval} for the details).

The WLS estimator weights the $i$-th observed value ($y_i$) of a random variable ($Y_i$)
with a weight $w_i$ in the estimation, expecting the weight to make the variance of $Y_i$ uniform.
Therefore, the weight is usually chosen to be reciprocal to the variance, i.e. $w_i=\sigma_{Y_i}^{-2}$.
We set $w_i$ to $\sigma_{\bYi}^{-2}$ with the variance model $\sigma_{\bYi}^{2}$ defined in Section~\ref{sec:var-model}.
Consequently,
the weight matrix $W$ becomes a function of $\bm{m}$, $\bm{n}$ and $\bm{k}$
with the variance model parameters $(q, \beta)$.
Note that it may be explicitly denoted by $W(\bm{m}, \bm{n}, \bm{k}; q, \beta)$, and
thus, $H$ by $H(\bm{m}, \bm{n}, \bm{k}; \hat{p}, \hat{a}, q, \beta)$.

Our goal is finding the RB configuration that minimizes the right-hand side of \eqref{eqn:ci},
so the objective function to be minimized is defined as
\begin{equation} \label{eqn:cii}
h(\bm{m}, \bm{n}, \bm{k}) \equiv
t_{M-3, 1-\alpha/2} \,
\sqrt{
H'(\bm{m}, \bm{n}, \bm{k}; \hat{p}, q, \beta)
}
\end{equation}
by omitting the constant factors on the right-hand side of \eqref{eqn:ci},
where $H'(\bm{m}, \bm{n}, \bm{k}; \hat{p}, q, \beta) = \hat{a}^2 H(\bm{m}, \bm{n}, \bm{k}; \hat{p}, \hat{a}, q, \beta)$.
The replacement of $H$ by $H'$ is only for factoring out an ineffective parameter $\hat{a}$.
Provided a confidence level $\alpha$ and some prior estimates of parameters $(\hat{p}, q, \beta)$ in advance,
the objective function $h$ becomes a function depending only on $ (\bm{m}, \bm{n}, \bm{k})$.
Note that the value of parameter $\hat{p}$ that we use to define the objective function
may differ from the actual decay rate estimated by an RB experiment we will run afterwards.

\begin{figure*}[tbp]
  \centering
  \begin{subfigure}[b]{0.36\textwidth}
      \includegraphics[width=\textwidth]{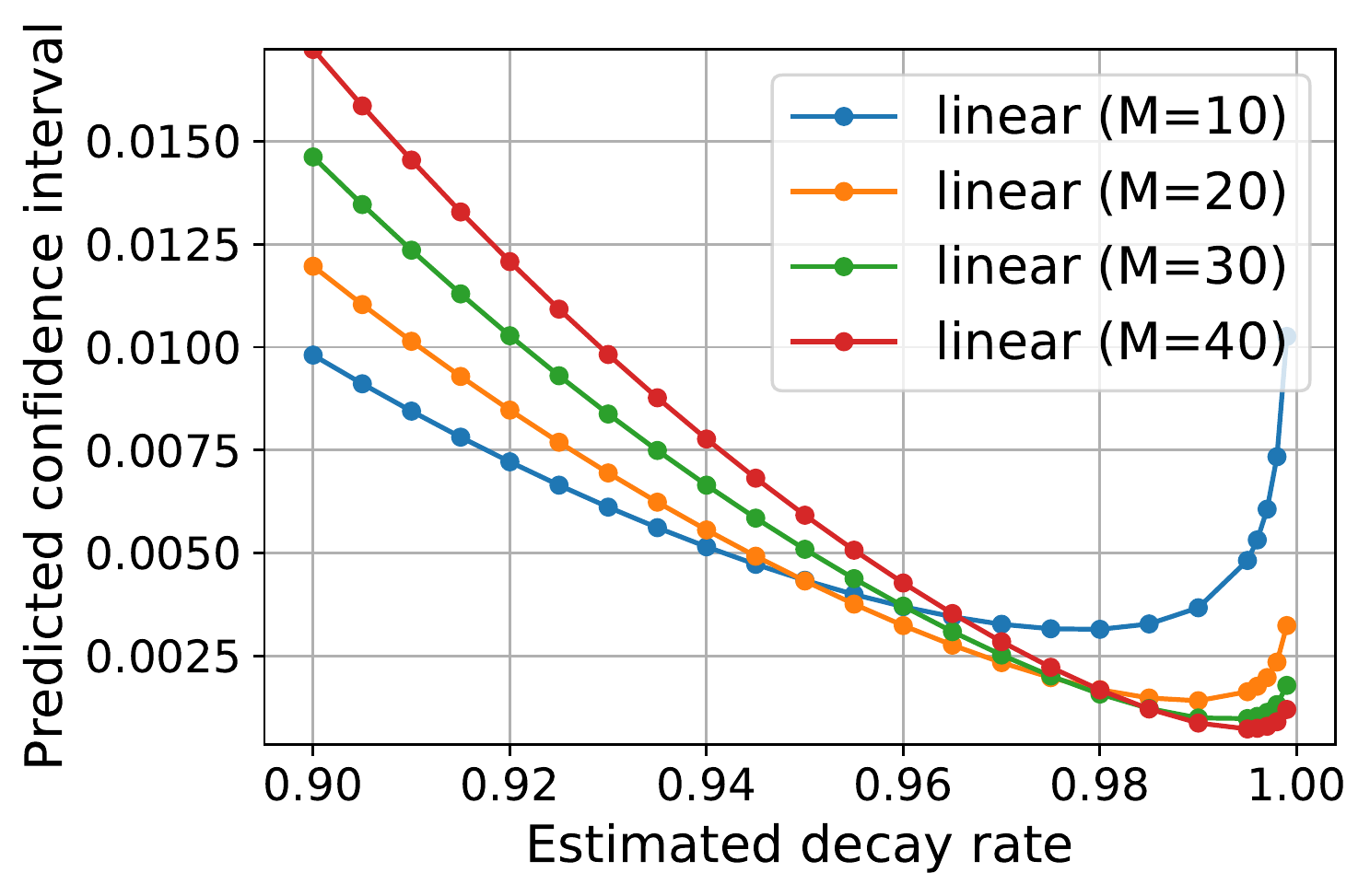}
      \caption{Linear configuration}\label{fig:linear}
  \end{subfigure}
  \begin{subfigure}[b]{0.3\textwidth}
      \includegraphics[width=\textwidth]{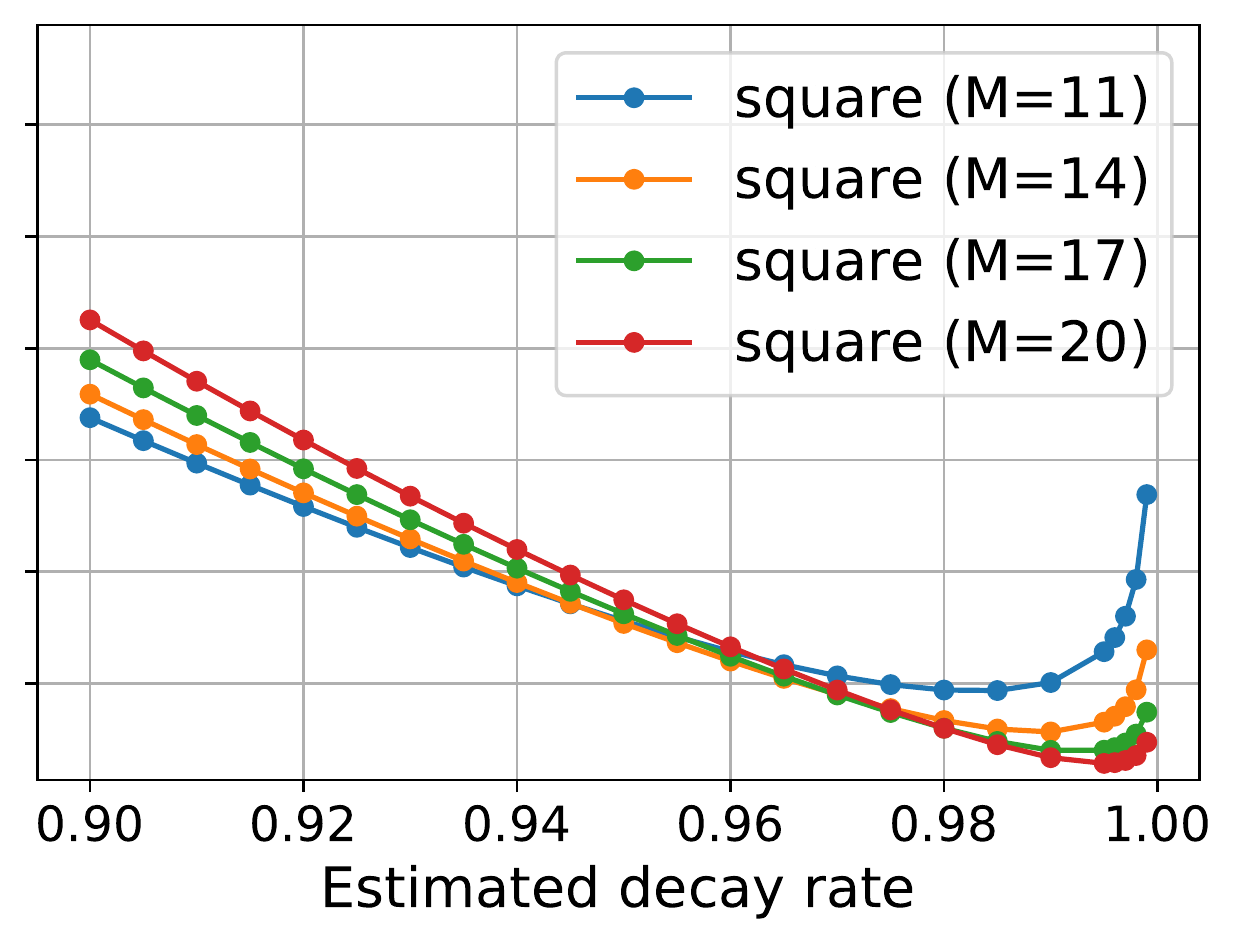}
      \caption{Square configuration}\label{fig:square}
  \end{subfigure}
  \begin{subfigure}[b]{0.3\textwidth}
      \includegraphics[width=\textwidth]{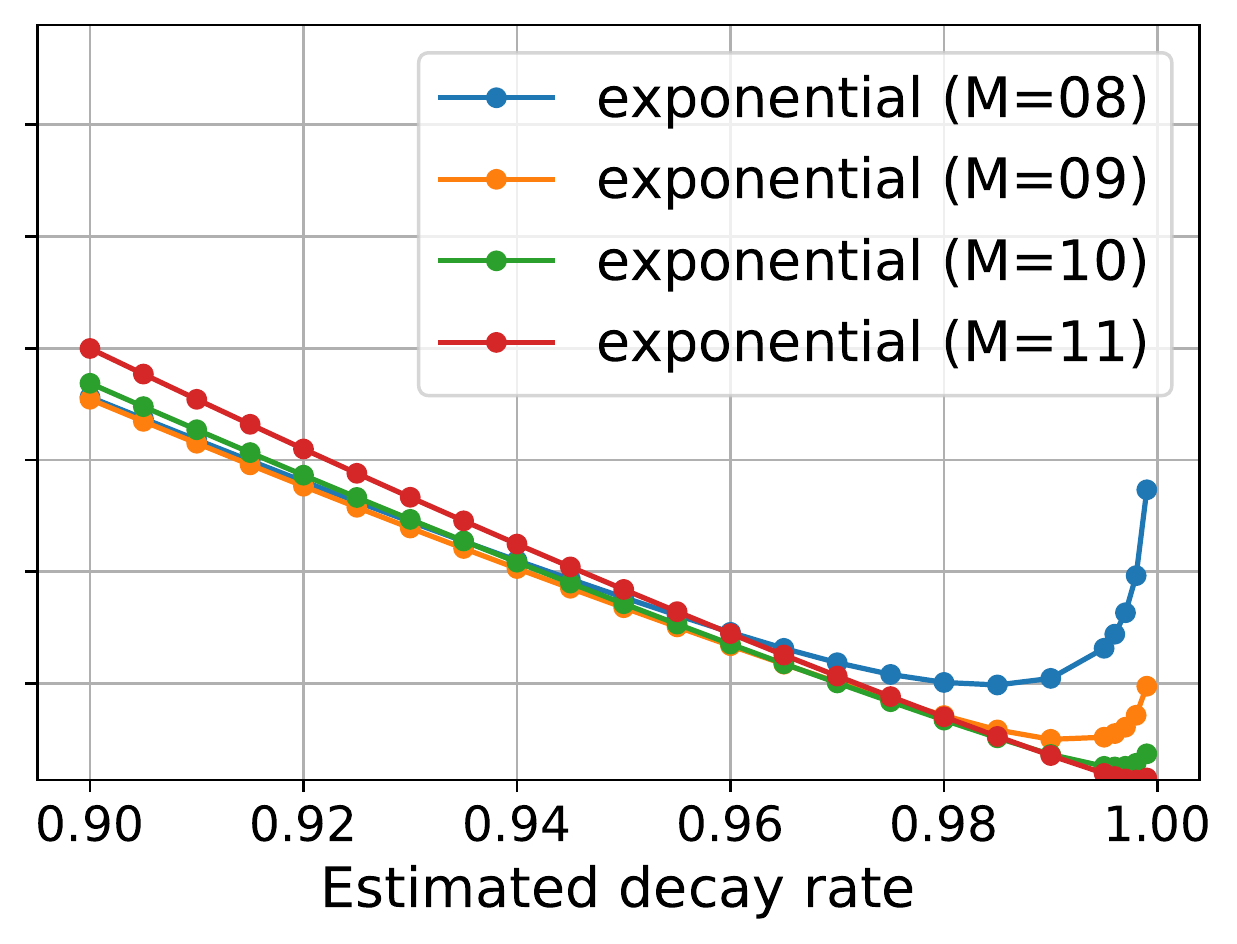}
      \caption{Exponential configuration}\label{fig:exponential}
  \end{subfigure}
  \caption{Confidence interval of estimated decay rate, predicted by the proposed method with different parameters (estimated decay rate and the dimension of Clifford lengths $M$) for each heuristic configuration}\label{fig:simulation}
\end{figure*}

\subsection{Execution Time Model} \label{sec:time-model}

We consider a simple approximated execution time model.
It estimates the time required for the execution of RB sequences
with a configuration $(\bm{m}, \bm{n}, \bm{k})$ by
\begin{equation}\label{eqn:approx-runtime}
t(\bm{m}, \bm{n}, \bm{k})
\approx
\sum_{i=1}^{M}{n_i \, k_i \, (c_1 \, m_i + c_0)}.
\end{equation}
Here, $c_1$ is a coefficient that reflects how execution time increases with Clifford length $m_i$,
and $c_0$ is a constant time required for running a sequence independent of Clifford length $m_i$.
In this model, some overhead time such as the loading time of control instructions due to sequence switching,
is not taken into account
(see Appendix~\ref{app:time-model} for approximation details).
The typical choice for $c_1$ is the average single Clifford gate time,
and that for $c_0$ is the sum of the average measurement time and the interval time between shots.
%

Given the time limit $T$ for the execution of an RB experiment,
we optimize the objective function under the time constraint
\begin{equation}\label{eqn:time-constraint}
t(\bm{m}, \bm{n}, \bm{k}) \leq T.
\end{equation}

\subsection{Formulation of Configuration Optimization} \label{sec:formulation}
Now we can formulate the problem of finding an optimal RB configuration
as a mathematical optimization problem as follows.
\begin{equation} \label{eqn:formulation}
    \begin{array}{ll}
  \mbox{minimize} & h(\bm{m}, \bm{n}, \bm{k}) \\
  \mbox{subject to}
        & m_{i} \geq m_{i-1} + 1,\ \mbox{for } i = 2, 3, \ldots, M,\\
        & t(\bm{m}, \bm{n}, \bm{k}) \leq T \\
        & \bm{m}, \bm{n}, \bm{k} \in \mathbb{Z}_+^M\ \mbox{(positive integer vector)}.\  \\
    \end{array}
\end{equation}
Here $h(\bm{m}, \bm{n}, \bm{k})$ is the confidence interval of the estimated decay rate defined in~\eqnref{eqn:cii},
$t(\bm{m}, \bm{n}, \bm{k})$ is the approximate execution time described in~\eqnref{eqn:approx-runtime},
and $T$ is the available time budget.
We introduce the constraint $m_{i} \geq m_{i-1} + 1$ to remove the symmetry in the formulation and make the problem more easily solvable.

There are two major difficulties in solving the (nonlinear) optimization problem \eqref{eqn:formulation} in practice:
\begin{enumerate}
\item The variability of the dimension of the Clifford lengths ($M$ may vary):
Most of optimization algorithms assume the dimension of solution space is fixed.
\item Integer variables ($\bm{m}, \bm{n}, \bm{k} \in \mathbb{Z}_+^M$):
Discrete solution spaces usually prevent the use of descent methods that performs well in practice.
\end{enumerate}

We overcome the aforementioned difficulties as follows.
\begin{enumerate}
\item We repeatedly solve the problem with a fix $M$ for $M=4, 5, \ldots, M_{\max}$,
and select the best among them.
We do not need to take such a large $M_{\max}$ in practice (typically, $M_{\max} \leq 40$),
so computation time is not a concern in solving multiple problem instances.
\item We relax the integer constraints and consider continuous variables
(i.e., $\bm{m}, \bm{n}, \bm{k} \in \mathbb{R}_+^M$).
We round the optimal solution for the relaxed problem and obtain a near-optimal integer solution for the original problem. 
Rounding continous solutions to discrete ones is standard to obtain approximate solutions of (mixed) integer programming~\cite{bertsimas2005optimization}. 
\end{enumerate}

\bigskip
\section{Experiments} \label{sec:experiment}
We conducted two experiments:
simulating the expected performance of typical heuristic configurations and
evaluating the effect of configuration optimization on real devices.
The first computational experiment was to show that
our proposed method can be useful to estimate the performance of a given configuration.
The second experiment was to determine if
our method can provide an optimal configuration that yields estimated decay rates
with small variance in a real environment.
Throughout both experiments, we investigated 2-qubit RB ($D=4$).
We implemented our method in Python and
used Qiskit 0.23.5, which is an open-source quantum computing software development framework~\cite{qiskit},
to generate and run sequences.
We used IBM Quantum systems~\cite{ibmquantum} with 5 qubits for experiments on real devices.
Hereafter, we omit the \textsf{ibmq\_} prefix in the device name for simplicity.

\begin{table*}[tbp]
 \centering
 \caption{Summary of RB configurations used in the experiment on real devices}
 \label{tab:configurations}
\begin{tabular}{lcccrr}
\hline
Configuration & $M$$^\dag$ & $\bm{m}$ & $\bm{n}$  & $N$$^\ddag$ & Estimated exec. time [s]  \\
\hline
\textsf{linear} & 21 & $[1, 11, 21, 31, \ldots, 201]$ & 5 (identical) & 105 & 3.261 \\
\textsf{square} & 17 & $[1, 4, 9, 16, \ldots, 289]$ & 6 (identical) & 102 & 3.193 \\
\textsf{exponential} & 10 & $[1, 2, 4, 8, \ldots, 512]$ & 10 (identical) & 100 & 3.114  \\
\textsf{optimal} & 16 & optimized & optimized & 99 & 2.974 \\
\textsf{optimal-identical-n} & 33 & optimized & 3 (identical) & 99 & 2.961  \\ \hline
\end{tabular}
\\
\vspace{1mm}
\footnotesize{
$^\dag$~The dimension of Clifford lengths $\bm{m}$,
$^\ddag$~The total number of sequences, i.e. sum of $\bm{n}$
}
\end{table*}

\subsection{Computational Experiment} \label{sec:exp-simu}
Our method can be useful to predict the confidence interval of a decay rate estimated by RB under a given configuration.
That also means it can simulate how the change in decay rate affects the performance
(i.e., predicted confidence interval) of a given configuration.
We conducted such a simulation for the three types of heuristic RB configurations:
\textsf{linear}, \textsf{square}, and \textsf{exponential}.
Each configuration designed to have its own Clifford lengths $\bm{m}$,
a number of sequences $n$ identical to the Clifford lengths,
and a fixed $k=100$.
Clifford lengths of a \textsf{linear}, \textsf{square} and \textsf{exponential} configuration were set to
$[10(x-1)+1]$, $[x^2]$ and $[2^{x-1}]$ for $x=1, \ldots, M$, respectively.
In the simulation, we fixed the confidence level $\alpha$ to 0.05.
The two parameters $(q, \beta)$ in the objective function~\eqnref{eqn:ci}
were fixed as $q=\hat{p}$ and $\beta=0.0025$.
The value of $\beta$ was obtained from a preliminary experiment on an IBM Quantum system \textsf{athens}.
The two parameters $(c_1, c_0)$ in the execution time model~\eqnref{eqn:approx-runtime}
were set as $c_1=0.6\,[\mu s]$ and $c_0=250\,[\mu s]$,
based on the public properties of \textsf{athens}:
The gate time of CNOT gate is around $0.4\,\mu s$
(one Clifford is composed of 1.5 CNOT gates on average),
the default intervals between shots is $250\,\mu s$, and
the measurement time is around $3\,\mu s$.
We set the time budget $T$ to 3 seconds.

Figure~\ref{fig:simulation} shows how the predicted confidence interval of the estimated decay rate
varies when changing the estimated decay rate $\hat{p}$ and the dimension of Clifford lengths $M$
for each heuristic configuration.
As expected, the optimal $M$ varied depending on the decay rate for all types of configurations.
Among the three types, \textsf{exponential} was most stable against the changing decay rate,
demonstrating its usefulness when there is little knowledge on the decay rate in advance.
It can be also seen that \textsf{linear} and \textsf{square} works well in a specific range of decay rates
if we can select the optimal $M$
when \textsf{square} is more stable than \textsf{linear}.
That suggests \textsf{square} is a good option if we have a good prior estimated decay rate.

In this way, our method can be used to predict the performance of heuristic configurations.
Moreover, it can be used to select the best among candidates of configurations.
For example, let consider any of the above heuristic configurations parameterized by $M$.
The confidence interval for each $M$ can be computed by fixing the decay rate to a given prior estimate.
Hence, the optimal $M$ that minimizes the confidence interval can be easily found, e.g., by grid search.
Actually, heuristic configurations compared with an optimized configuration in the next section are prepared this way.

\begin{table*}[tbp]
 \centering
 \caption{Adjusted standard deviation (raw standard deviation) of estimated decay rates by multiple runs of RBs on different real devices with different configurations.
The best value among the configurations is in bold for each device.
 Average estimated decay rates over runs are stated in Avg $\hat{p}$ column for reference.}
 \label{tab:comparison}
\begin{tabular}{lcr|ccc|cc}
\hline
\multicolumn{3}{}{} & \multicolumn{3}{|c|}{Heuristic configurations} & \multicolumn{2}{c}{Optimized configurations (Proposed)} \\
Device & Avg $\hat{p}$ & Runs & \textsf{linear} & \textsf{square} & \textsf{exponential} & \textsf{optimal} & \textsf{optimal-identical-n} \\
\hline
\textsf{athens} & 0.9706 & 98 & 0.001481 (0.002843) & 0.001217 (0.002658) & 0.001275 (0.002721) & \textbf{0.001074} (0.002798) & 0.001151 (0.002720) \\
\textsf{quito} & 0.9731 & 96 & 0.001469 (0.004673) & 0.001268 (0.004245) &0.001378 (0.004419) & \textbf{0.000967} (0.004357) & 0.001190 (0.004310) \\
\textsf{bogota} & 0.9762 & 96 & 0.001160 (0.002483) & 0.001107 (0.002507) & 0.001231 (0.002648) & \textbf{0.000860} (0.002529) & 0.000971 (0.002397) \\
\textsf{rome} & 0.9836 & 93 & 0.000886 (0.001017) & \textbf{0.000774} (0.000953) & 0.000914 (0.001088) & 0.000911 (0.000919) & 0.000849 (0.001089) \\
\textsf{lima} & 0.9858 & 92 & 0.000979 (0.001201) & 0.000767 (0.000934) & \textbf{0.000730} (0.000910) & 0.000844 (0.001064) & 0.000798 (0.000956) \\ \hline
\end{tabular}
\end{table*}

\subsection{Experiment on Real Devices} \label{sec:exp-real}
The optimal configuration found by our method may deviate from the true optimal configuration
value due to the following two gaps.
One is the modeling gap, i.e., the models our method relies on cannot perfectly represent the real system.
The other is the gap between the prior estimates and true values of the model parameters.
These gaps are inevitable and too difficult to measure in practice.
Therefore, we aim to demonstrate that our method is still useful in practice
even after subtracting the impact of those gaps.
That is, the configuration optimized by our method can yield the estimated decay rate with smaller variance than those from heuristic configurations in RBs on real devices.

Therefore, we conducted the experiment as follows.
\begin{enumerate}
\item Preparing five configurations in total to be compared:
Three heuristic configurations with optimized $M$
(\textsf{linear}, \textsf{square}, \textsf{exponential}; the same as discussed in Section~\ref{sec:exp-simu}),
a configuration optimized by our method (\textsf{optimal}), and
a reference configuration optimized by our method with restricting the number of sequences to be identical (\textsf{optimal-identical-n}).
\item Bundling five sets of sequences generated with the five configurations up into one job.
We run 100 jobs on each of the five devices (\textsf{athens}, \textsf{quito}, \textsf{bogota}, \textsf{rome} and \textsf{lima}), which corresponds to running 5 configurations $\times$ 5 devices $\times$ 100 jobs $=$ 2,500 RBs.
\item Analyzing the deviation in the decay rates estimated via 100 times of RBs for each device and  configuration.
\end{enumerate}
We ran RBs with common configurations on the five devices,
each of which represents a different extent of imperfect prior estimates of model parameters.
By comparing the results from different devices,
we investigated how the imperfections in model parameters affect the performance of each configuration
(i.e. the standard deviation of the resulting estimated decay rates).
%
Throughout this experiment, we focused on optimizing $\bm{m}$ and $\bm{n}$,
and we fixed the number of shots to 100 ($k=100$).

In the first preparation step,
we used the same values as those used in the Section~\ref{sec:exp-simu}
for the model parameters required in our method:
$\alpha=0.05$, $\hat{p}=q=0.97$, $\beta=0.0025$, $c_1=0.6\,[\mu s]$, $c_0=250\,[\mu s]$, and $T=3\,[s]$.
Note that the parameters $\hat{p}$, $q$, and $\beta$ are determined by a preliminary experiment on \textsf{athens}.
We used \texttt{scipy.optimize} module to optimize the relaxed problem discussed in Section~\ref{sec:formulation}
when computing \textsf{optimal} and \textsf{optimal-identical-n}
(see Appendix~\ref{app:nonconvex} for the details).
%
%
We set $M_{\max}$ to 40 for both configurations.
The computation time to optimize each configuration was within 30 seconds (for all values of $M$) on a laptop PC with an Intel Core i7 2.7 GHz and 16 GB memory.

Table~\ref{tab:configurations} summarizes five configurations prepared in the first step.
For the optimized configuration (\textsf{optimal}),
$\bm{m}=$[1, 2, 19, 21, 23, 24, 25, 26, 27, 28, 29, 51, 52, 105, 195, 369],
$\bm{n}=$[8, 5, 5, 5, 6, 6, 5, 6, 6, 7, 5, 5, 5, 5, 8, 12].
For the optimized configuration with the identical $n$ constraint (\textsf{optimal-identical-n}),
$\bm{m}=$[1, 2, 3, 4, 5, 12, 20, 23, 24, 25, 26, 27, 28, 29, 30, 31, 32, 33, 34, 35, 36, 37, 38, 39, 53, 92, 136, 181, 227, 276, 329, 385, 445],
$n=3$.
Note that we allowed to exceed the time budget constraint (due to rounding) for heuristic configurations
while the optimal ones strictly comply with the constraint.
Thus, it ensures that the comparison does not favor the optimal ones (i.e. our method) unfairly.
It is interesting to note that \textsf{optimal-identical-n} seems to sample a similar number of neighbor Clifford lengths for each Clifford length that is heavily sampled in \textsf{optimal}; therefore, they appear to be sampled from the same distribution.

In the second running step,
we ran the sequences on five IBM Quantum devices with five qubits available to us as of May 19, 2021\footnote{
We tried to run on all six 5-qubit devices, but \textsf{belem} was too busy at that time and failed to complete all jobs within three days, so we removed it from the analysis.
}.
We ran 100 jobs for each of the five devices, and a few of them failed while waiting in the job queue for unknown reasons.
We analyzed the results of 98, 96, 96, 93, and 92 successful jobs from
\textsf{athens}, \textsf{quito}, \textsf{bogota}, \textsf{rome} and \textsf{lima}, respectively.
We used qubit 0 and 1 for all the devices.

Table~\ref{tab:comparison} shows the adjusted standard deviations of decay rates estimated by the WLS estimator from the survival rate data obtained by repeated RBs on the five real devices.
Because it took about 24 hours to run all jobs for each device (including the wait time in the queue),
the raw values of standard deviations described in $(\cdot)$ in the table
were affected by the temporal variation in the real decay rate.
This can be observed from the changes of the average estimated decay rate over five configurations for each job
at \textsf{athens} as shown in Fig.~\ref{fig:temporal-athens}
(see Appendix~\ref{app:temporal} for similar figures of other devices).
\begin{figure}[htbp]
\centering
\includegraphics[width=.95\columnwidth]{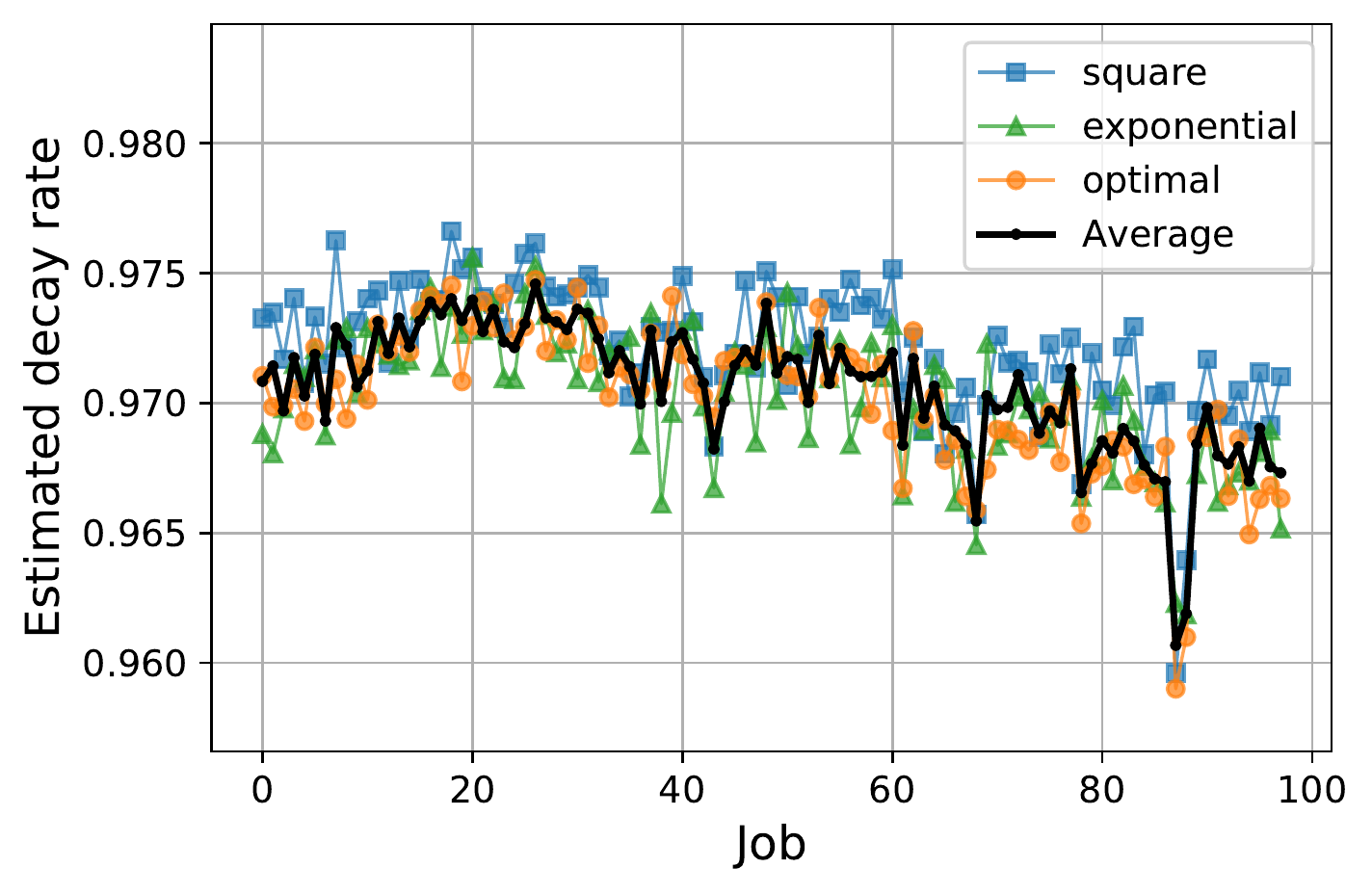}
\caption{Temporal variation in decay rate at \textsf{athens}: the black line depicts the average of estimated decay rate over all five configurations}
\label{fig:temporal-athens}
\end{figure}
Therefore, the values adjusted to remove the effect of time fluctuation were described in the table and used for the evaluation.
Specifically, the mean value of the five configurations minus the bias for each configuration was used as the expected value of the configuration at the job execution.
The \emph{adjusted standard deviation} was defined as the sample deviation calculated by the residual from the expected value.
The bias for each configuration was set to the difference between the average of decay rates over all jobs and configuration, 
and the average of decay rates over all jobs for each configuration.

As shown in the table,
\textsf{optimal} achieves the least deviation of the estimated decay rate by RBs on \textsf{athens}, \textsf{quito}, and \textsf{bogota}.
This is the expected result because we used the values based on \textsf{athens}
as the prior estimates for the model parameters.
In particular, we set the prior estimate of the decay rate to 0.97,
while the actual (estimated) decay rate of \textsf{athens}, \textsf{quito}, and \textsf{bogota} were
all close to this value at 0.9706, 0.9731, and 0.9762, respectively.
The improvement rates from \textsf{square} (the best of the heuristic configurations) to \textsf{optimal} in the standard deviation were quite large at
11.7\% for \textsf{athens}, 23.7\% for \textsf{quito}, and 22.3\% for \textsf{bogota}.
In contrast, heuristic configurations achieved slightly less deviation than the optimal configuration
in \textsf{rome} and \textsf{lima}, where the actual decay rates (0.9836 and 0.9858) were larger than its prior estimate (0.97).
This may be because the errors in prior estimates were so large that the optimization did not work as expected.
Supporting the results on real devices, \textsf{optimal} achieves better results on noisy simulators of real devices: either the best or the second best (see Appendix~\ref{app:simulation} for the details of noisy simulation).
It is interesting to note that \textsf{optimal-identical-n} likely provided more stable results
against the errors in prior estimates of decay rate comparing with \textsf{optimal}.
This could be because the identical-n constraint mitigated the risk of over-optimization along with the given model parameters
as if the regularization term prevents overfitting in machine learning.

In summary, our method is capable of finding a better configuration than typical heuristic configurations, at least when we have sufficiently accurate prior estimates of parameters required in the method.

\medskip
\section{Discussion}
In this paper, we developed a method for minimizing the confidence interval under the time budget constraint.
Our method can easily be modified so to minimize the execution time while bounding the confidence interval by replacing the formulation in~\eqnref{eqn:formulation} with
\begin{equation}
    \begin{array}{ll}
  \mbox{minimize} & t(\bm{m}, \bm{n}, \bm{k}) \\
  \mbox{subject to}
        & m_{i} \geq m_{i-1} + 1,\ \mbox{for } i = 2, 3, \ldots, M,\\
        & h(\bm{m}, \bm{n}, \bm{k}) \leq \epsilon \\
        & \bm{m}, \bm{n}, \bm{k} \in \mathbb{Z}_+^M\ \mbox{(positive integer vector)}.\  \\
    \end{array}
\end{equation}
where $\epsilon$ is the feasible upper bound of the confidence interval.

An advantage of our method is that it can be used for the first RB of a brand new device
because the confidence model used in our method only requires the prior estimate of parameters, not their learning with training data.
In the subsequent RBs, we could predict a more accurate confidence interval
by using the decay rate estimated in the previous RB as the prior estimate.
Ultimately, we could obtain the sample data from every RB, which can be used to improve the confidence model by learning from the data.
In that case, such approaches that use machine learning techniques (as discussed in~\cite{granade2015accelerated}) would work well.

In this paper, we focused on the standard RB protocol and
proposed a method for optimizing a sampling strategy for it.
Investigating how our method can be extended to other benchmarking protocols with random sampling,
such as interleaved randomized benchmarking~\cite{magesan2012efficient},
dihedral benchmarking~\cite{carignan2015characterizing,cross2016scalable,harper2017estimating},
or other variants~\cite{helsen2019multiqubit,proctor2019direct},
is left for future work.

Furthermore, our method may be generalized to be combined with
other protocols or algorithms that have the following properties:
\begin{itemize}
\item the protocol has a fitting parameter to be estimated in it (decay rate of RB),
\item the protocol requires a fitting model used in it (survival rate decay model of RB), and
\item the protocol repeats experiments to obtain samples to be fitted (survival rates of RB)
with changing sampling parameters (RB configuration).
\end{itemize}
%
For example, an amplitude estimation algorithm proposed in~\cite{tanaka2020amplitude}
satisfies the above properties by mapping
the fitting parameter to an angle related to the amplitude to be estimated,
the fitting model to Equation~(13) in the paper, and
the sampling parameter to the number of repetitions of the amplifying operation.

\section{Conclusion}
We addressed the problem of optimizing sampling strategies for the standard RB.
We showed how a sampling strategy is determined by configurable parameters (an RB configuration):
Clifford lengths, the number of sequences for each Clifford length, and the number of shots.
We discussed how the variance of the survival rate may not be uniform with respect to Clifford length
and discussed how to model the heteroskedasticity in a simple form.
We proposed a method for optimizing an RB configuration,
which constructs a mathematical optimization problem
that minimizes the confidence interval of the estimated decay rate
while keeping the predicted execution time within a given time budget.
The method does not change the RB protocol itself, so it is easily utilized as preprocessing.
Our experiment on real devices demonstrated that the proposed method can find a better configuration,
i.e. achieving smaller deviation in the resulting estimated decay rate,
than typical heuristic configurations in practice.
We believe the proposed method would be useful in practice
to reduce the sampling error while maintaining the execution time,
or to reduce the execution time while maintaining the sampling error.

\section*{Acknowledgment}
We thank
Ikko Hamamura,
Naoki Kanazawa,
Antonio C\'{o}rcoles,
Shelly Garion, and
Chris J. Wood
for their helpful comments.

\bibliographystyle{IEEEtran}
\bibliography{rb-plan-opt}

\appendix

\subsection{Approximation in the Variance Model of Survival Rate} \label{app:var-model}
We first explain more about the approximation in the variance model described in~\eqnref{eqn:varYmodel}--\eqnref{eqn:varshot}.
We suppose that the mean survival rate of the $j$-th random sequence with Clifford length $m_i$,
denoted by $\mu_{ij}$,
follows some distribution whose mean is $\mu_i \equiv a \, p^{m_i} + b$ and variance is $\stdseq^2$.
We also suppose that the survival count over $k_i$ shots for the $j$-th random sequence with Clifford length $m_i$ follows a binomial distribution Bi$(k_i, \mu_{ij})$.
Let $Y_{ij}$ denote the survival rate over $k_i$ shots for the $j$-th sequence with Clifford length $m_i$.
The survival rate $Y_{ij}$ approximately follows a normal distribution $\mathcal{N}(\mu_{ij}, \sigma_{ij}^2)$
for large $k_i$ where $\sigma_{ij}^2 = \frac{\mu_{ij}(1 - \mu_{ij})}{k_i}$.
Note that we are interested in the average of the average survival rate
$\bar{Y_i} \equiv \frac{\sum_{j=1}^{n_i}{Y_{ij}}}{n_i}$
and its variance $\sigma_{\bar{Y_i}}^2$.

Assuming the shot sampling error is independent of the circuit sampling error,
we obtain \eqnref{eqn:varYmodel} where
\begin{equation}
\stdshot^2
= \frac{\frac{1}{n_i}\sum_{j=1}^{n_i}{\mu_{ij}(1 - \mu_{ij})}}{k_i}.
\end{equation}
Because $\mu_{ij}$ is difficult to know in advance, we need to approximate $\stdshot^2$ more.
We assume $\mu_{ij}$ to be identical and replace $\mu_{ij}$ with $\mu_{i}$
and thus obtain \eqnref{eqn:varshot}.


\begin{figure*}[htbp]
  \centering
  \begin{subfigure}[b]{0.24\textwidth}
      \includegraphics[width=\textwidth]{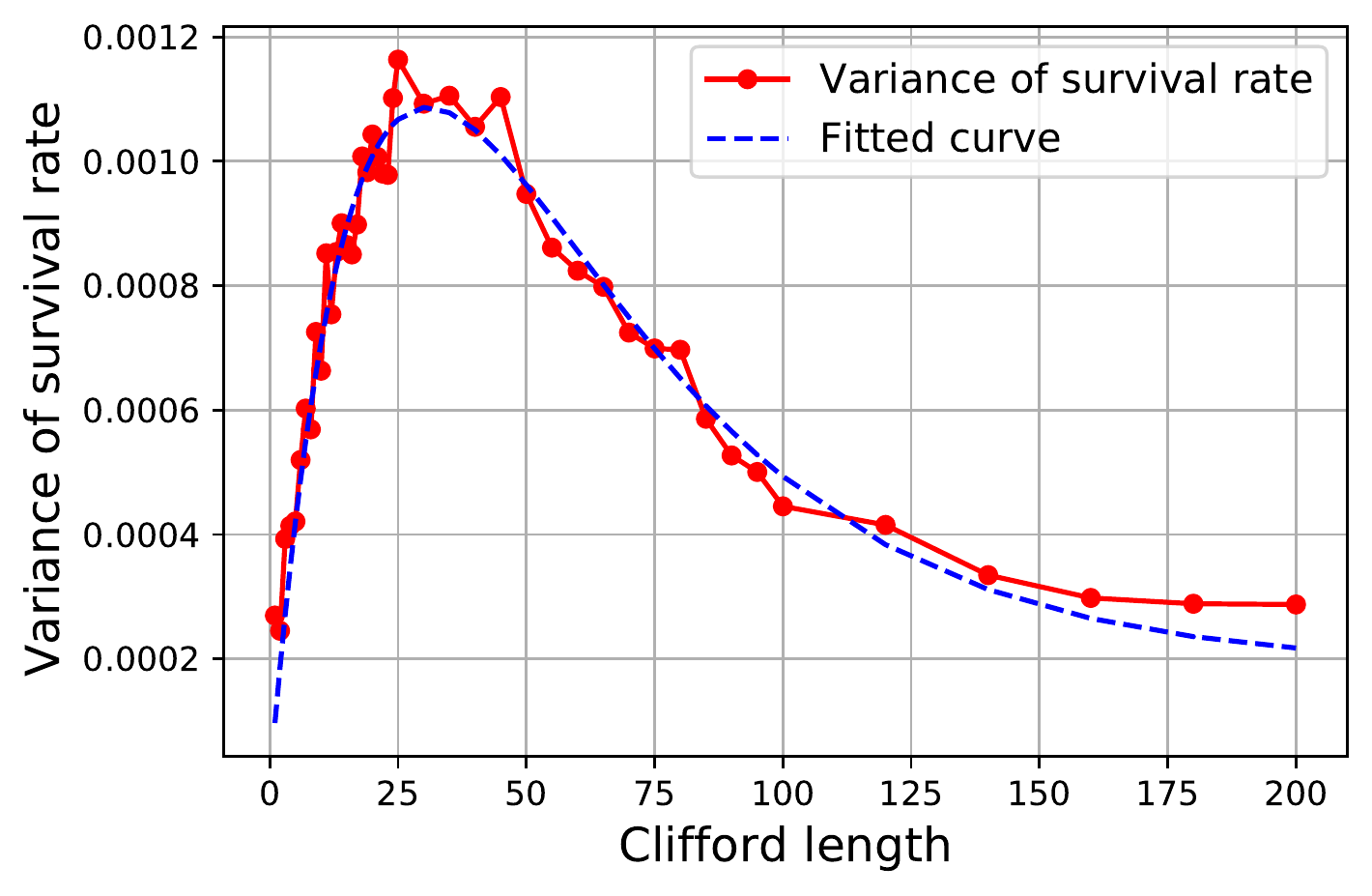}
      \caption{\textsf{quito}}\label{fig:var-quito}
  \end{subfigure}
  \begin{subfigure}[b]{0.24\textwidth}
      \includegraphics[width=\textwidth]{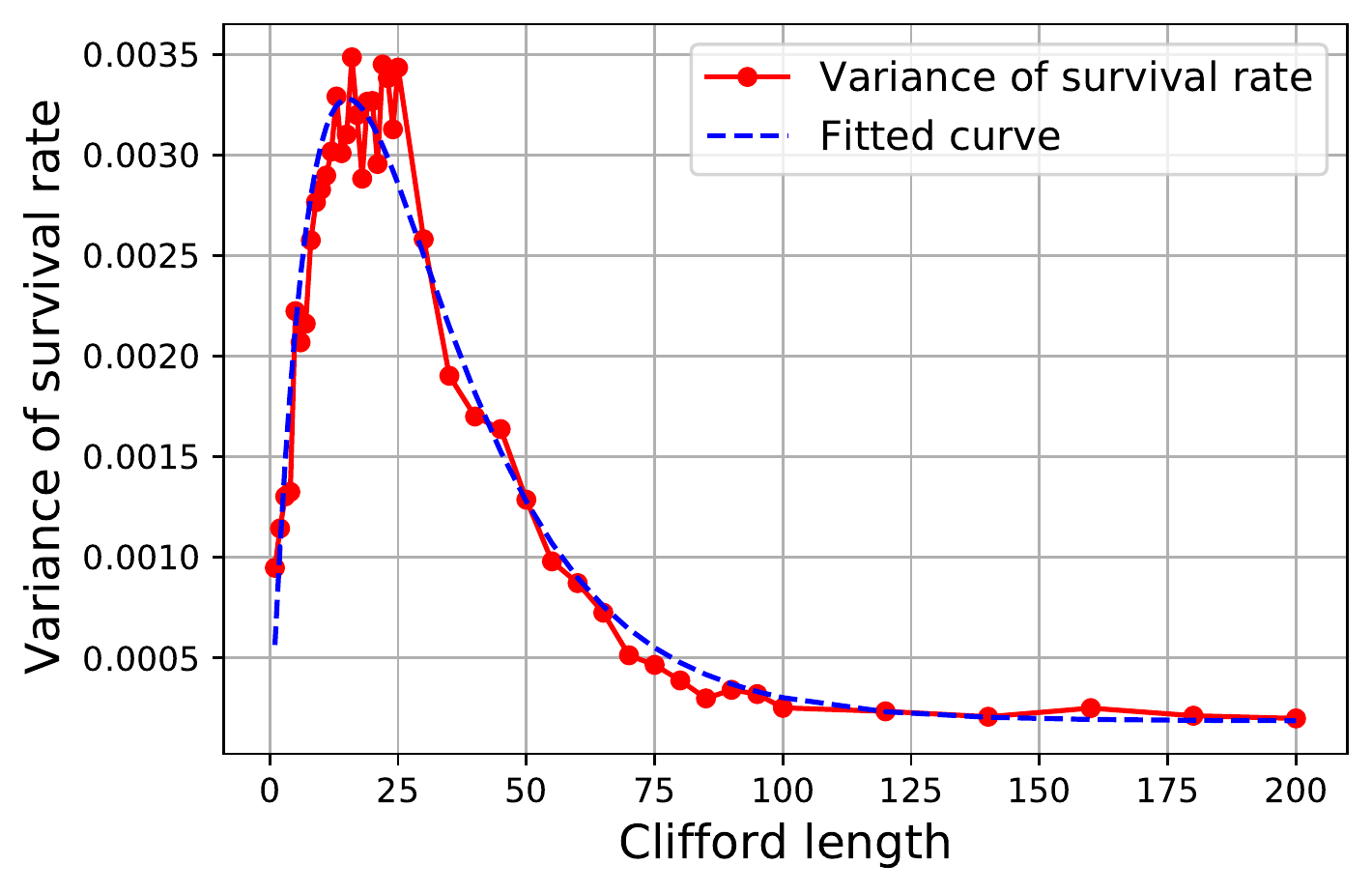}
      \caption{\textsf{bogota}}\label{fig:var-bogota}
  \end{subfigure}
  \begin{subfigure}[b]{0.24\textwidth}
      \includegraphics[width=\textwidth]{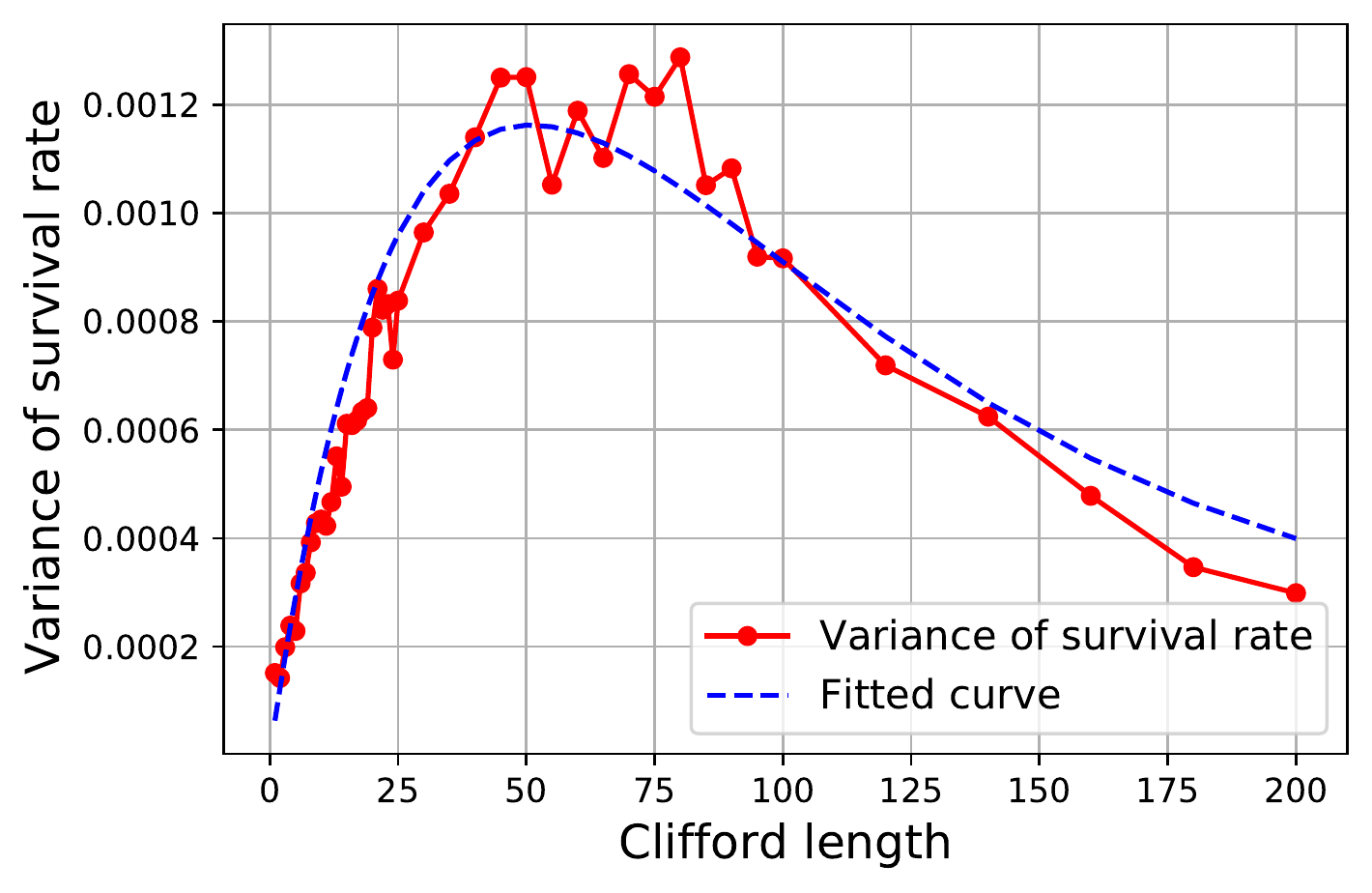}
      \caption{\textsf{rome}}\label{fig:var-rome}
  \end{subfigure}
  \begin{subfigure}[b]{0.24\textwidth}
      \includegraphics[width=\textwidth]{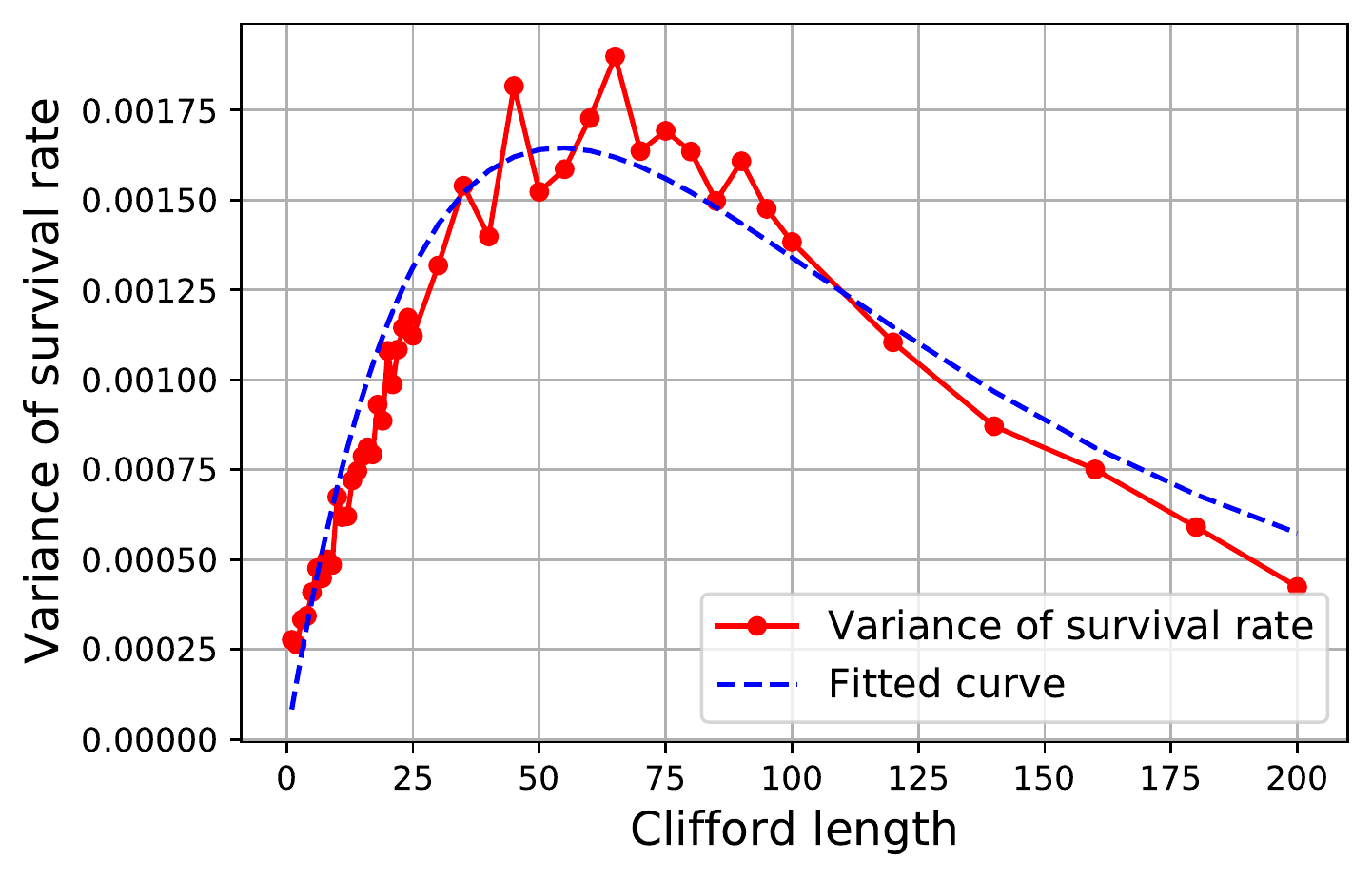}
      \caption{\textsf{lima}}\label{fig:var-lima}
  \end{subfigure}
  \caption{Sample variance of survival rate for each Clifford length on different devices and the fitted curve to the variance model}\label{fig:varYfits}
\end{figure*}

We provide four more sets of sample variance data obtained from the devices of IBM Quantum systems,
\textsf{quito}, \textsf{bogota}, \textsf{rome} and \textsf{lima},
and the fitted curve for each device in Fig.~\ref{fig:varYfits}.
In each figure, the sample variance of survival rate for each Clifford length is plotted as red points and
the blue dashed line shows the fitted curve to the model $n_i \, \sigma_{\bar{Y_i}}^2$
defined by~\eqnref{eqn:varYmodel}--\eqnref{eqn:varshot}.
As shown in the figure, the sample variance of survival rate is not uniform with respect to Clifford length for all devices.
The model relatively explains the sample variance well
although some divergence in the tail is observed in several cases.

\begin{figure*}[htbp]
  \centering
  \begin{subfigure}[b]{0.24\textwidth}
      \includegraphics[width=\textwidth]{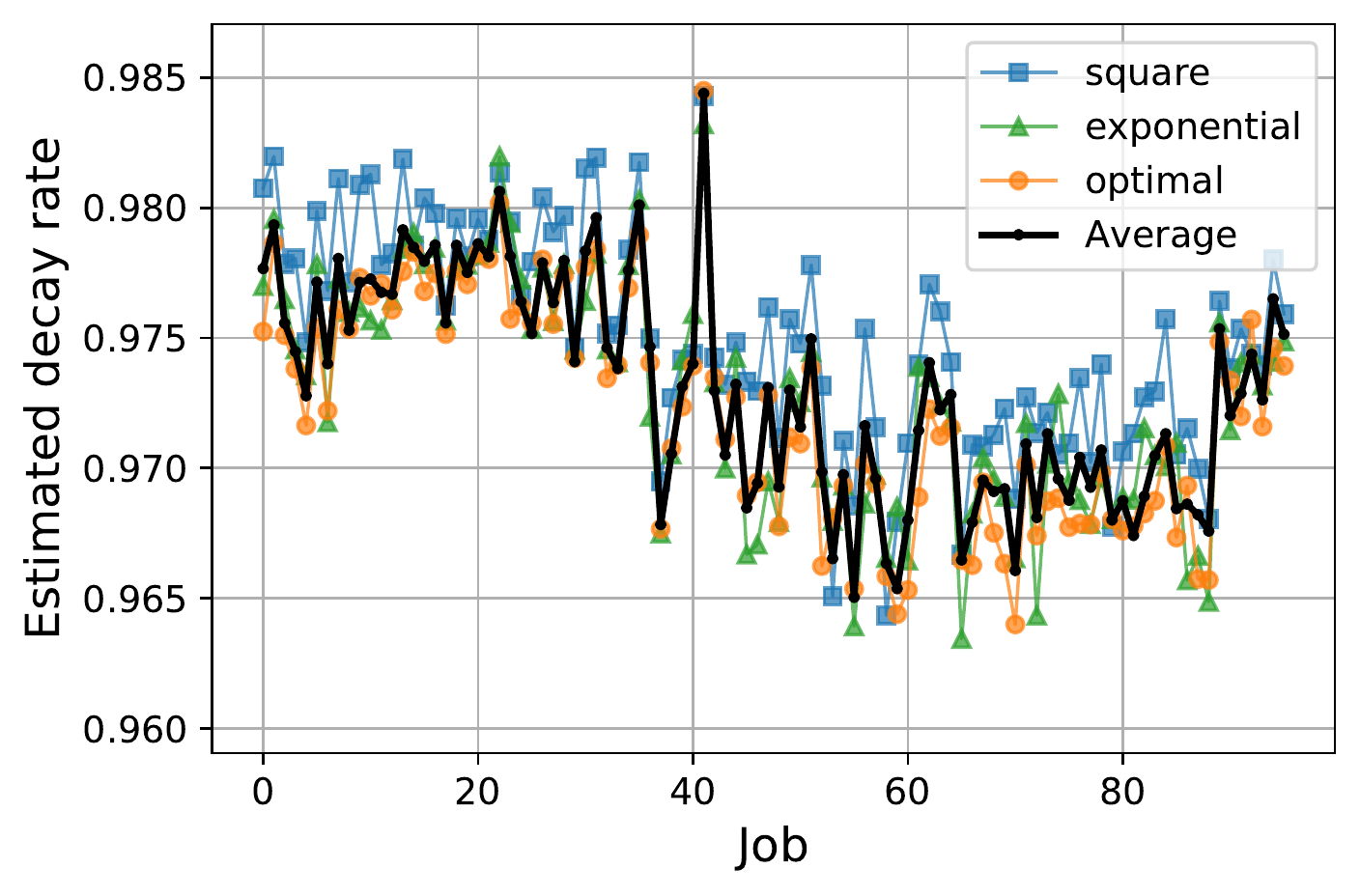}
      \caption{\textsf{quito}}\label{fig:temporal-quito}
  \end{subfigure}
  \begin{subfigure}[b]{0.24\textwidth}
      \includegraphics[width=\textwidth]{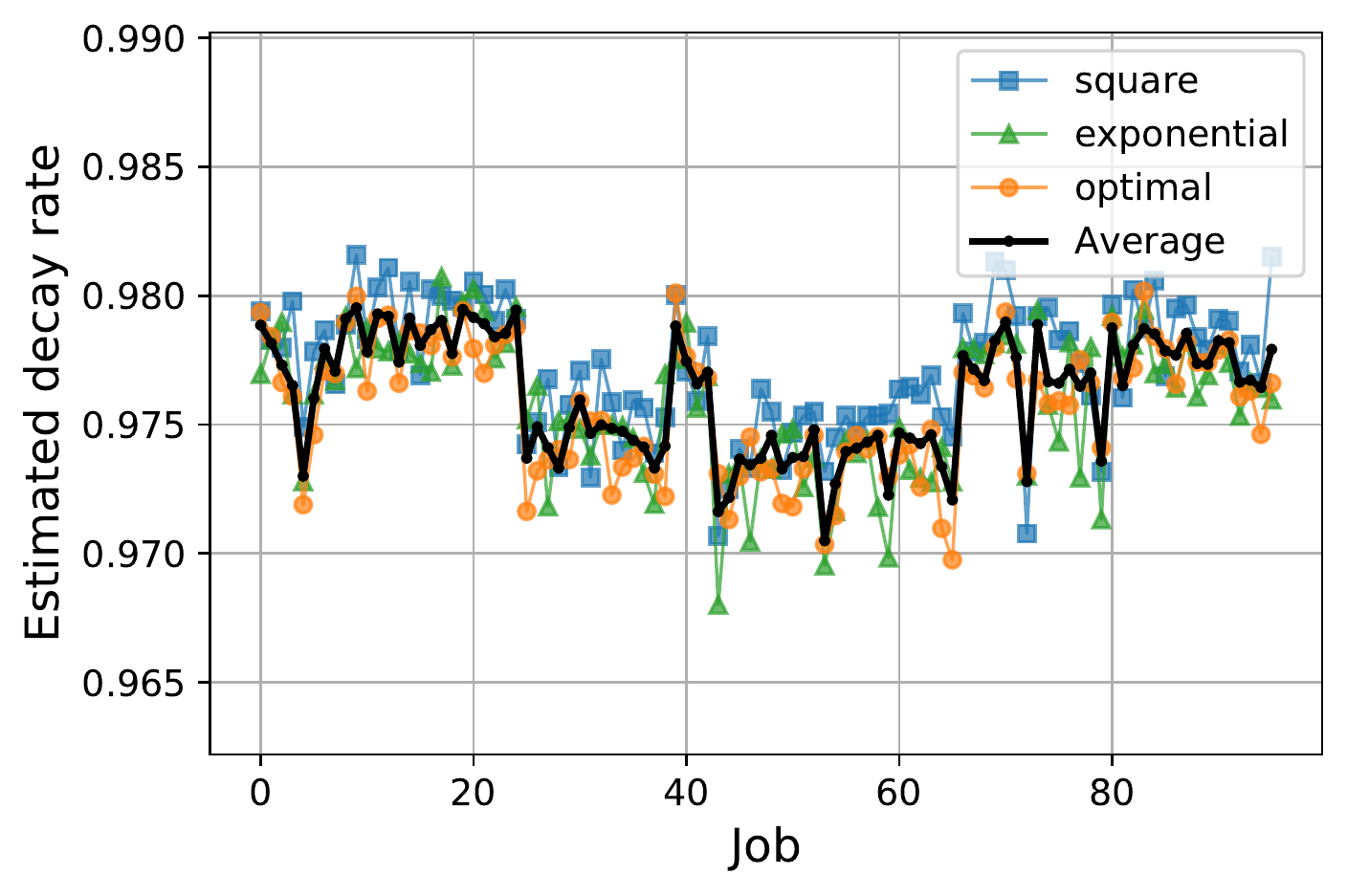}
      \caption{\textsf{bogota}}\label{fig:temporal-bogota}
  \end{subfigure}
  \begin{subfigure}[b]{0.24\textwidth}
      \includegraphics[width=\textwidth]{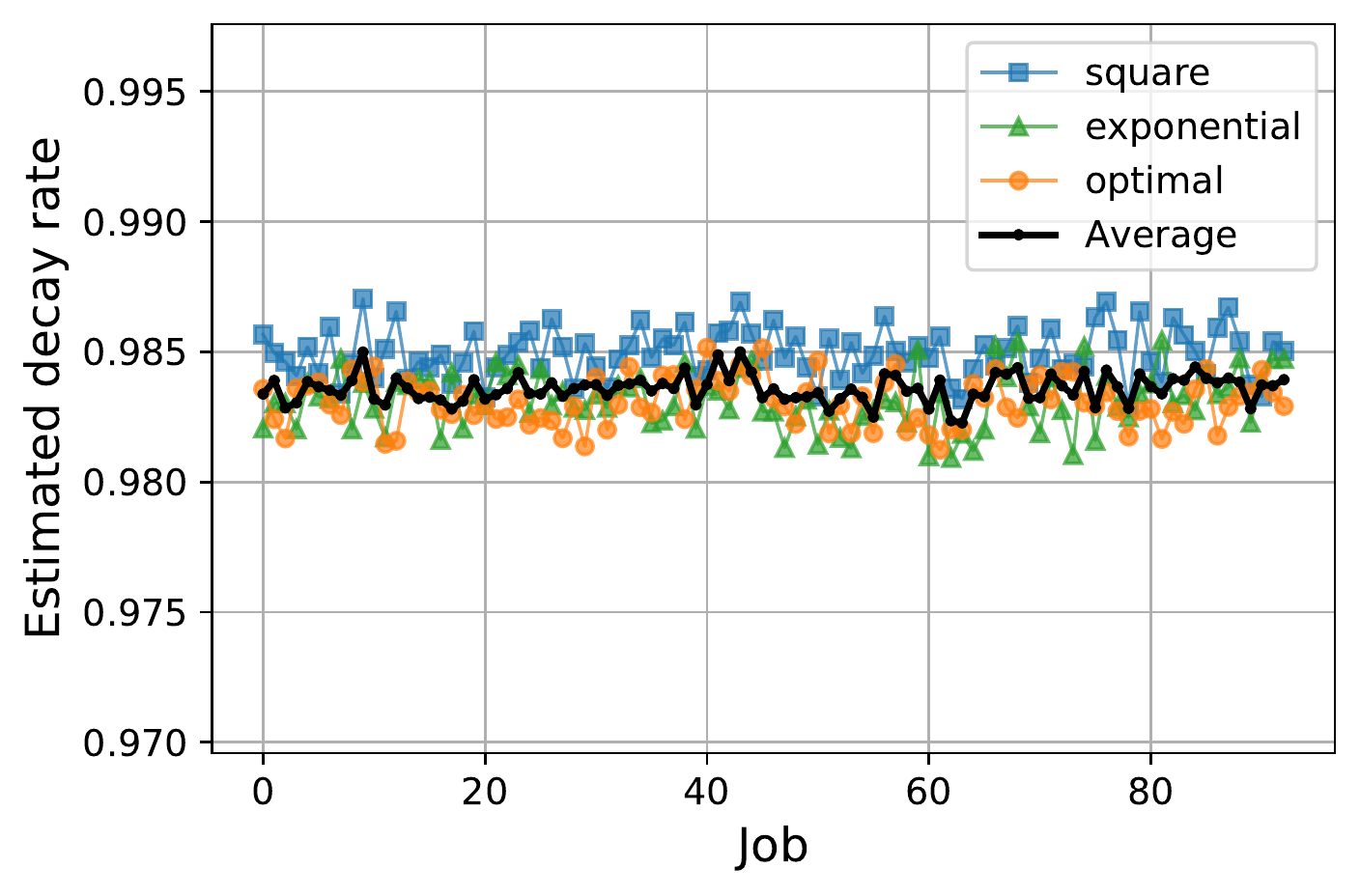}
      \caption{\textsf{rome}}\label{fig:temporal-rome}
  \end{subfigure}
  \begin{subfigure}[b]{0.24\textwidth}
      \includegraphics[width=\textwidth]{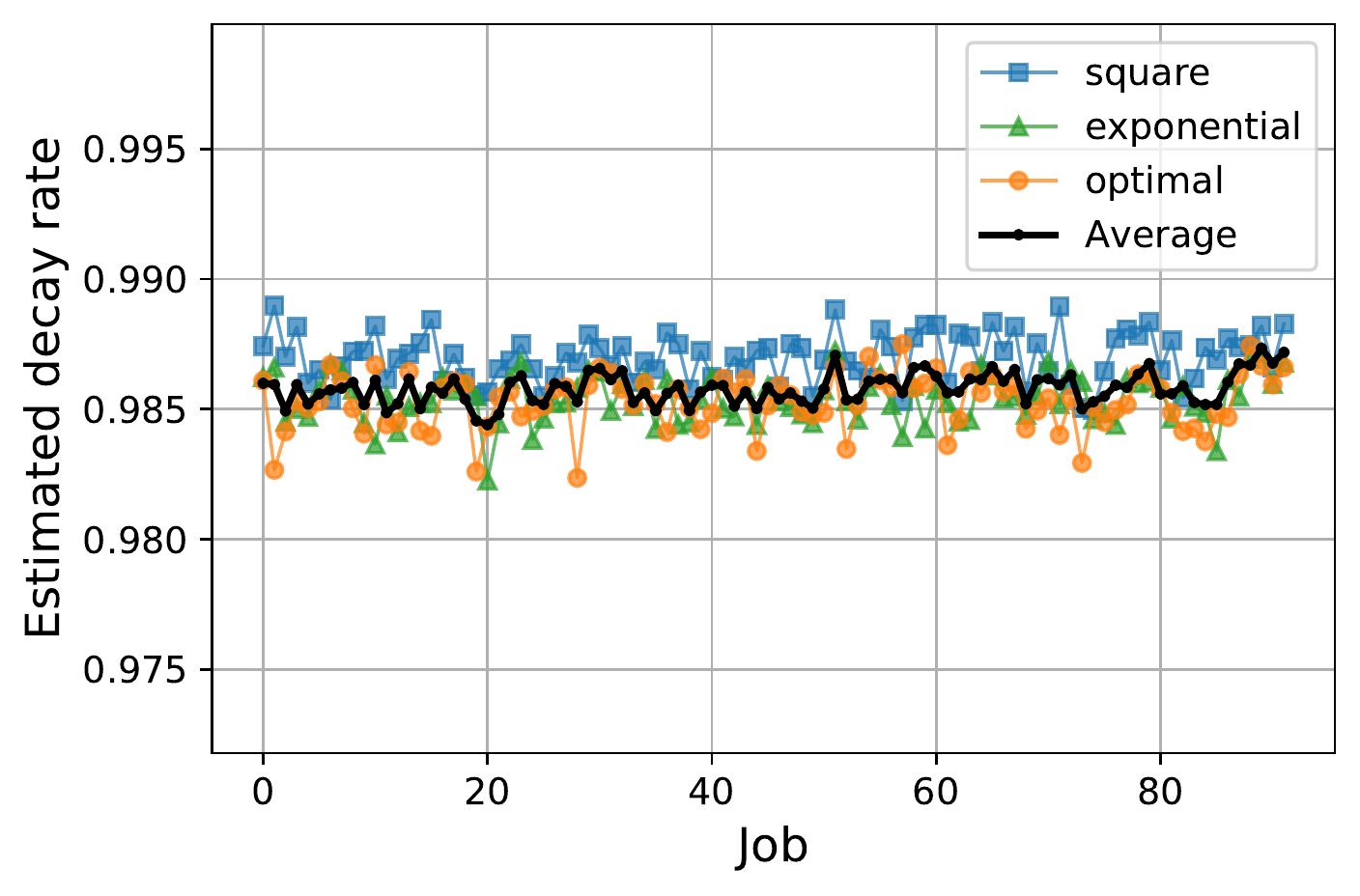}
      \caption{\textsf{lima}}\label{fig:temporal-lima}
  \end{subfigure}
  \caption{Temporal variation in decay rate on different devices: The average estimated decay rate over all five configurations by job (black line) and the estimated decay rate with \textsf{square}/\textsf{exponential}/\textsf{optimal} by job (blue/green/orange)}\label{fig:temporals}
\end{figure*}

\subsection{Explicit Form of the Confidence Interval Function} \label{app:conf-interval}
We provide the explicit expression of $H'$ used in the definition of confidence interval function discussed in Section~\ref{sec:conf-interval}.
Recall that
\begin{equation}
H' = \hat{a}^2 H = \hat{a}^2
\left[\left(J(\bm{\hat{\theta}})^T W J(\bm{\hat{\theta}}) \right)^{-1}\right]_{\hat{p}\hat{p}}
\end{equation}
and $J(\bm{\hat{\theta}})^T W J(\bm{\hat{\theta}})$ is a 3x3 matrix
in the form of
$$
\left[
\begin{array}{cc}
A & \bm{c} \\
\bm{c}^T & u
\end{array}
\right]
$$
where
$$
A = \left[
\begin{array}{cc}
a^2 \, \sum_i{w_i \, m_i^2 \, p^{2m_i-2}} &
a \, \sum_i{w_i \, m_i \, p^{2m_i-1}}  \\
a \, \sum_i{w_i \, m_i \, p^{2m_i-1}} &
\sum_i{w_i \, p^{2m_i}}
\end{array}
\right],
$$
$$
c^T = \left[
a \, \sum_i{w_i \, m_i \, p^{m_i-1}},\ 
\sum_i{w_i \, p^{m_i}}
\right],
$$
and $u = \sum_i{w_i}$.
Here and hereafter, we omit the superscript `$\hat{\ }$` of $p$,  $a$,  $b$ for brevity.
Using the block matrix inversion formula
$$
\left[
\begin{array}{cc}
A & \bm{c} \\
\bm{c}^T & u
\end{array}
\right]^{-1}
=
\left[
\begin{array}{cc}
(A - \frac{\bm{c} \, \bm{c}^T}{u})^{-1} & * \\
* & *
\end{array}
\right]
$$
and the explicit form of the inverted 2x2 matrix, we obtain
\begin{widetext}
\begin{equation} \label{eqn:Hprime}
H' \equiv
\frac{\left[ \sum{w_i \, p^{2m_i}} - \frac{(\sum{w_i \, p^{m_i}})^2}{u} \right]}{
\left[ \sum{w_i \, p^{2m_i}} - \frac{(\sum{w_i \, p^{m_i}})^2}{u} \right]
\left[ \sum{w_i \, m_i^2 \, p^{2m_i-2}} - \frac{(\sum{w_i \, m_i \, p^{m_i-1}})^2}{u} \right]
-
\left[ \sum{w_i \, m_i \, p^{2m_i-1}} - \frac{(\sum{w_i \, p^{m_i}})(\sum_i{w_i \, m_i \, p^{m_i-1}})}{u} \right]^2
}.
\end{equation}
\end{widetext}

\subsection{Approximation in the Execution Time Model} \label{app:time-model}
We explain how the approximate execution time model described in~\eqnref{eqn:approx-runtime} can be derived.
We start from a general execution time model, that is
the time required for the execution of a RB
with a configuration $(\bm{m}, \bm{n}, \bm{k})$
is estimated by
\begin{equation}\label{eqn:general-runtime}
t(\bm{m}, \bm{n}, \bm{k})
\equiv
\sum_{i=1}^{M}{n_i \left\{L(m_i) + k_i \, R(m_i)\right\}}
\end{equation}
where $L(m_i)$ represents the loading time of a sequence with Clifford length $m_i$ and
$R(m_i)$ represents the execution time of a sequence with Cliffords length $m_i$.
We focus on the execution time on a QPU (quantum processing unit) and
omit the compiling time of sequences.
Assuming that the loading time is constant and the execution time is linear in $m_i$,
i.e. $L(m_i)=c_L$ and $R(m_i)=c_1 \, m_i + c_0$,
\begin{equation}
t(\bm{m}, \bm{n}, \bm{k})
\approx
\sum_{i=1}^{M}{n_i \left\{c_L + k_i \, (c_1 \, m_i + c_0)\right\}}.
\end{equation}
In addition, assuming the loading of sequences is pipelined and its time is negligible,
i.e. $c_L=0$, the model follows \eqnref{eqn:approx-runtime}.

The model given in~\eqnref{eqn:general-runtime} assumes that
$k_i$ shots of a common sequence are done consecutively.
However, some systems may repeat the single shot of $n_i$ sequences $k_i$ times.
In that case, the model could be given as
\begin{equation}\label{eqn:approx-runtime-2}
t(\bm{m}, \bm{n}, \bm{k})
\equiv
\sum_{i=1}^{M}{n_i \, k_i \left\{L(m_i) + R(m_i)\right\}}
\end{equation}
\\*
Fortunately, if applying the same approximation to this model,
it results in the same model as \eqnref{eqn:approx-runtime}.

\vspace{10mm}
\subsection{Solving the Optimization Problem} \label{app:nonconvex}
The optimization problem~\eqnref{eqn:formulation} is obviously nonlinear in $\bm{m}$ and $\bm{n}$,
recalling that the objective function includes the term $\sqrt{H'}$ with $H'$ written down in~\eqnref{eqn:Hprime}
and $w_i$ is proportional to $\sqrt{n_i}$ because $w_i$ is set to $\sigma_{\bYi}^{-2}$.
Actually, it is even non-convex.
In general, the non-convex optimization problem has multiple locally optimal solutions and
it is often difficult to find the globally optimal solution.
In the experiments in Section~\ref{sec:exp-real},
we used \texttt{scipy.optimize.minimize} function
(using default parameters except for \texttt{tol}=1.0e-10)
to compute one of the locally optimal solutions.
We added extra constraints $n_i \geq 5$ in order to mitigate the impact of outliers in the formulation for \textsf{optimal}.
We replaced the vector variable $\bm{n}$ with a scalar variable $n$ to represent a common number of sequences in the formulation for \textsf{optimal-identical-n}.
We set the initial guess (\texttt{x0}) to $\bm{m}=[1, 2, 3, \ldots, M]$ and
$n_i=5$ for all $i$ (for \textsf{optimal}) and $n=3$ (for \textsf{optimal-identical-n}).
Notice that we may find a better solution by changing the initial guess.

\subsection{Temporal Variation in Decay Rate} \label{app:temporal}
We provide four more plots of the temporal variation in decay rate on IBM Quantum devices,
\textsf{quito}, \textsf{bogota}, \textsf{rome} and \textsf{lima} in Fig.~\ref{fig:temporals}.
The estimated decay rate with \textsf{square}, \textsf{exponential} and \textsf{optimal} for each job are also shown in the figure 
(we omit that with \textsf{linear} and \textsf{optimal-identical-n} for ease of reading).
As seen in the figures, \textsf{square} was likely to estimate slightly higher decay rates than the other configurations for all devices.
In fact, the sample average of decay rates from \textsf{square} was
0.9750, 0.9772, 0.9837, and 0.9870
while those from others were within
[0.9723, 0.9728], [0.9757, 0.9762], [0.9825, 0.9828], and [0.9852, 0.9857]
for \textsf{quito}, \textsf{bogota}, \textsf{rome} and \textsf{lima}, respectively.
This might imply that we need more investigation on the bias in the estimated decay rate.

\subsection{Results of Noisy Simulation} \label{app:simulation}
To confirm RB configurations optimized by our method performs as expected in a synthetic environment,
we conducted the same experiments described in Section~\ref{sec:exp-real}
on noisy simulators.
We utilized \texttt{NoiseModel.from\_backend} function in Qiskit to simulate noisy execution of sequences on each device.
We show the results in Table~\ref{tab:simulation},
which confirm that \textsf{optimal} can achieve better results on noisy simulators of real devices:
either the best or the second best.

\begin{table*}[tbp]
 \centering
  \caption{Standard deviations of estimated decay rates by 100 runs of RBs with different configurations on noisy simulators of different real devices.
The best value among the configurations is in bold for each device.
Average estimated decay rates over runs are stated in Avg $\hat{p}$ column for reference.}
 \label{tab:simulation}
\begin{tabular}{lcr|ccc|cc}
\hline
\multicolumn{3}{}{} & \multicolumn{3}{|c|}{Heuristic configurations} & \multicolumn{2}{c}{Optimized configurations (Proposed)} \\
Device & Avg $\hat{p}$ & Runs
& \textsf{linear} & \textsf{square} & \textsf{exponential}
& \textsf{optimal} & \textsf{optimal-identical-n} \\
\hline
\textsf{athens} & 0.9758 & 100 & 0.001311 & 0.001125 & 0.001277 & \bf{0.001032} & 0.001104 \\
\textsf{quito} & 0.9779 & 100 & 0.001188 & 0.001052 & 0.001342 & 0.001028 & \bf{0.001024} \\
\textsf{bogota} & 0.9805 & 100 & 0.000927 & 0.000910 & 0.000909 & 0.000860 & \bf{0.000846} \\
\textsf{rome} & 0.9829 & 100 & 0.000888 & 0.000941 & 0.000992 & 0.000838 & \bf{0.000778} \\
\textsf{lima} & 0.9825 & 100 & 0.001111 & 0.000946 & 0.000975 & 0.000843 & \bf{0.000792} \\ \hline
\end{tabular}
\end{table*}

\end{document}